\documentclass[aps,prb,twocolumn,floatfix,superscriptaddress]{revtex4-2}
\usepackage{mathrsfs}
\usepackage{graphicx}
\usepackage[english]{babel}
\usepackage{amsmath}
\usepackage{amssymb}
\usepackage{xcolor}
\usepackage{relsize}
\usepackage{CJK}
\usepackage{dsfont}
\usepackage{bm}% bold math

\newcommand{\me}{\mathrm{e}}
\newcommand{\mi}{\mathrm{i}}

\newcommand{\dif}{\mathrm{d}}

\allowdisplaybreaks
\begin{document}
\title{Theory of the Uhlmann Phase in Quasi-Hermitian Quantum Systems}
\author{Xu-Yang Hou}
\affiliation{School of Physics, Southeast University, Jiulonghu Campus, Nanjing 211189, China}
\author{Xin Wang}
\affiliation{School of Physics, Southeast University, Jiulonghu Campus, Nanjing 211189, China}
\author{Hao Guo}
\email{guohao.ph@seu.edu.cn}
\affiliation{School of Physics, Southeast University, Jiulonghu Campus, Nanjing 211189, China}
\affiliation{Hefei National Laboratory, University of Science and Technology of China, Hefei 230088, China}
\begin{abstract}
Geometric phases play a fundamental role in understanding the geometric structure of quantum states, yet extending the Uhlmann phase to non-Hermitian systems poses significant challenges due to parameter-dependent inner product structures. In this work, we develop a comprehensive theory of the Uhlmann phase for quasi-Hermitian systems, where the physical Hilbert space metric varies with external parameters. By constructing a generalized purification that respects the quasi-Hermitian inner product, we derive the corresponding parallel transport condition and Uhlmann connection. Our analysis reveals that the parameter-dependent metric modifies the Uhlmann connection and leads to a finite-temperature distribution of Uhlmann-phase regions that differs from the standard Hermitian case. Applying this formalism to solvable two-level models, we uncover tunable finite-temperature Uhlmann-phase diagrams, where the parameter-dependent metric shifts and deforms the zeros of the Uhlmann amplitude, thereby reshaping the temperature intervals in which nontrivial Uhlmann phases occur. Furthermore, by extending established interferometric protocols originally developed for Hermitian systems, the geometric amplitude can be recast as a measurable Loschmidt amplitude between purified states, providing a practical and experimentally accessible pathway to investigate quasi-Hermitian mixed-state geometric phases and their finite-temperature transitions. This work establishes a unified framework for understanding mixed-state geometric phases in quasi-Hermitian quantum systems and, through its natural relation to the mixed-state quantum geometric tensor, opens new avenues for exploring local geometry in quasi-Hermitian thermal states.
\end{abstract}
\maketitle
\section{Introduction}
The geometric phase, since its modern formulation by Berry, has proven to be a fundamental concept permeating diverse areas of quantum physics, from the dynamics of adiabatic systems to the characterization of topological phases of matter~\cite{Pancharatnam1956, Berry1984}. While the Berry phase provides a robust framework for pure quantum states, realistic physical systems are invariably open, interacting with their environment and existing at finite temperature. This has motivated a long-standing quest to generalize geometric phases to mixed quantum states, described by density matrices. Several approaches have been developed in this direction, including the interferometric geometric phase proposed by Sj\"oqvist et al.~\cite{PhysRevLett.85.2845}, which has been extensively studied~\cite{PhysRevA.67.020101, PhysRevLett.90.160402, Faria_2003} and observed in various experimental platforms such as nuclear magnetic resonance~\cite{PhysRevLett.91.100403}, polarized neutrons~\cite{PhysRevLett.101.150404}, and Mach-Zehnder interferometry~\cite{PhysRevLett.94.050401}. Another important mathematical generalization, formulated independently by Uhlmann, provides a fiber bundle description in which a geometric phase is accumulated via the parallel transport of the amplitude of a density matrix~\cite{Uhlmann1986, Uhlmann1989, Uhlmann1991, UHLMANN1995461}. This approach, now known as the Uhlmann phase, has gained considerable attention for its deep geometric structure and its relevance to condensed matter and quantum information~\cite{Viyuela14, ViyuelaPRL14-2, ourPRB20, OurPRB20b, Galindo21, Zhang21,PhysRevA.104.023303, Hou2023}. This phase has proven instrumental in exploring the finite-temperature geometry of condensed matter systems~\cite{Viyuela14, ViyuelaPRL14-2, PhysRevLett.113.076407, PhysRevA.98.033816, PhysRevA.104.042204, Galindo21, PhysRevB.110.134319, 98sq-16bz, prq8-c9ns}.

Historically, these developments were primarily formulated within the framework of Hermitian quantum mechanics. Meanwhile, a vast and rapidly growing class of physical systems is described by non-Hermitian Hamiltonians, which have uncovered many fascinating phenomena including Anderson localization~\cite{PhysRevLett.77.570}, unconventional behavior of quantum emitters~\cite{GZ22, Roccati_2022}, and distinctive topological properties~\cite{PhysRevX.8.031079, PhysRevLett.121.026808, Roccati2023}. Theoretical and experimental investigations of non-Hermitian topology have since flourished, revealing unique phenomena such as non-Hermitian skin effects and novel bulk-boundary correspondences~\cite{PhysRevLett.121.086803, Li2019NatCommun, PhysRevLett.123.230401, doi:10.1126/science.aaw8205, PhysRevLett.126.083604, Xiao2020NatPhys, PhysRevLett.125.126402}. Most recently, this pursuit has uncovered a hidden (2+1)-dimensional electrodynamics in pseudo-Hermitian two-level systems, where a topological instanton sources the Berry curvature and quantizes the magnetic flux jump~\cite{kcls-cn82}. The notion of geometric phase has been extended to non-Hermitian systems~\cite{JPAGW13}, and subsequently applied to construct the quantum geometric tensor for such systems~\cite{PhysRevA.99.042104}.  In parallel, Uhlmann-based mixed-state topology has also been explored in non-Hermitian systems through the Uhlmann phase and thermal Uhlmann-Chern numbers~\cite{Yang2026MixedStateTopology}.

Among non-Hermitian frameworks, pseudo-Hermitian and, more specifically, quasi-Hermitian systems stand out, as they possess real spectra and admit a consistent quantum mechanical interpretation through the construction of a physically admissible positive-definite metric operator $\eta_+$~\cite{RevModPhys.15.175, PhysRevLett.80.5243, Mostafazadeh2002I, Mostafazadeh2002II, Mostafazadeh2002III}. A major branch includes systems with parity-time reversal ($\mathcal{PT}$) symmetry, which has attracted considerable research attention~\cite{PhysRevLett.80.5243, PhysRevB.82.052404, PhysRevLett.110.083604, Korff_2007, Korff_2008, RevModPhys.88.035002, Bender1999} and has been experimentally realized across acoustics, optics, electronics, and quantum systems~\cite{Cham_2015, Feng_2017}. The dynamics and geometry of such systems are defined not by the standard Dirac inner product, but by a physical inner product $(\cdot,\cdot)_{\eta_+} = \langle \cdot | \eta_+ | \cdot \rangle$ that generally depends on the system's parameters~\cite{doi:10.1142/S0219887810004816, Das_2011}. This parameter-dependent inner product fundamentally alters the geometric structure of the Hilbert space, raising a critical question: how should the geometric phase of mixed states be understood in such a setting?

In this work, we address this question by developing a comprehensive theory of the Uhlmann phase for quasi-Hermitian quantum systems at finite temperature. Our motivation is twofold. First, the need to describe realistic systems inherently subject to environmental influences and thermal fluctuations naturally calls for a mixed-state geometric framework. Extending this framework to encompass the quasi-Hermitian paradigm is essential for a complete description of open or non-Hermitian systems in thermal equilibrium. Second, the geometric properties of mixed states in quasi-Hermitian systems are expected to differ profoundly from those of their pure-state counterparts. Unlike the Berry phase, which for a quasi-Hermitian system is governed by adiabatic evolution under a single Hamiltonian, the Uhlmann phase involves the parallel transport of a purification and is sensitive to the system's statistical mixture. The interplay between this statistical mixing and the parameter-dependent metric $\eta_+(\lambda)$ is anticipated to give rise to novel geometric features not captured by either the standard Uhlmann theory or the quasi-Hermitian Berry phase alone.

A central observation underlying our construction is that the standard purification $\rho = WW^\dagger$, used ubiquitously in conventional Uhlmann theory, is no longer valid for quasi-Hermitian systems, as it would implicitly force the density matrix to be Hermitian. To overcome this fundamental obstruction, we instead decompose the density matrix using the $\ddagger$-operation defined by the physical metric $\eta_+(\lambda)$: $\rho = W W^{\ddagger}$, where $W$ is the amplitude or purification. In this framework, the gauge redundancy is naturally described by the quasi-unitary group $U_{\eta_+}(N)$, whose elements satisfy $U^{\ddagger} U = \mathbb{I}$. Consequently, the amplitude can be expressed as $W = \sqrt{\rho}\,U$, with $U \in U_{\eta_+}(N)$ serving as the generalized phase factor that replaces the unitary factor of the Hermitian theory. This construction faithfully reflects the underlying metric structure and provides the proper starting point for generalizing the Uhlmann formalism to the quasi-Hermitian realm. By imposing a parallel transport condition based on minimizing the $\eta_+$-weighted Hilbert-Schmidt distance between neighboring amplitudes, we derive the corresponding Uhlmann connection $\mathcal{A}_{\text{U}}^{\eta}$ and the holonomy that defines the quasi-Hermitian Uhlmann phase $\theta_{\text{U}}^{\eta}$. We show that this geometric structure is uniquely determined by the parameter dependence of both the density matrix and the metric, thereby capturing the subtle interplay between statistical mixing and the evolving physical inner product.

To illustrate the physical implications of our formalism, we apply it to two exactly solvable two-level models. The first features a parameter-independent metric and serves as a baseline, while the second, a $\mathcal{PT}$-symmetric model, exhibits a genuinely parameter-dependent metric, allowing us to examine how the metric reshapes the finite-temperature Uhlmann-phase structure. Our analysis reveals that the parameter-dependent metric $\eta_+(\lambda)$ does not merely provide a static background; rather, it enters the Uhlmann connection and shifts the zero lines of the Uhlmann amplitude, thereby giving rise to tunable finite-temperature geometric phase transitions. As a result, the temperature intervals in which nontrivial Uhlmann phases occur are no longer determined solely by the winding number, but can be controlled by the quasi-Hermitian coefficient. Finally, we elucidate the relationship between our quasi-Hermitian construction and the standard Uhlmann theory by examining the effect of a parameter-dependent similarity transformation that maps the quasi-Hermitian system to a Hermitian representation. This analysis reveals that the two frameworks, while describing the same physics, are connected through a nontrivial, metric-induced gauge structure. The parameter dependence of the metric, encoded in $\dif\eta_+(\lambda)$, gives rise to an additional connection term that modifies the parallel transport rule for the purified states. This finding highlights that the local geometry of mixed states in quasi-Hermitian systems differs from its Hermitian counterpart through the parameter-dependent metric and is closely linked to the broader context of quantum geometric tensors for mixed states.

Our work establishes a unified geometric framework for mixed states in quasi-Hermitian systems, paving the way for exploring finite-temperature geometry in a wide range of non-Hermitian physical settings, from open quantum systems to correlated materials. The remainder of this paper is organized as follows. In Sec.~II, we review the quasi-Hermitian framework and establish the parameter-dependent metric structure. In Sec.~III, we construct the quasi-Hermitian Uhlmann formalism based on the $\eta_+$-weighted purification scheme and derive the corresponding connection and Uhlmann phase. In Sec.~IV, we discuss two solvable two-level examples with parameter-independent and parameter-dependent metrics, respectively, and examine the associated finite-temperature geometric phase transitions. In Sec.~V, we discuss an experimentally feasible interferometric protocol for measuring the quasi-Hermitian Uhlmann phase. Finally, Sec.~VI presents our conclusions and outlook.

\section{Mathematical Description of Quasi-Hermitian Mixed States}
\subsection{Definitions and Basic Properties of Quasi-Hermitian Systems}
In non-Hermitian quantum systems, a fundamental challenge is to establish a consistent probabilistic interpretation and geometric structure while preserving dynamical consistency. If a Hamiltonian is not Hermitian under the standard Dirac inner product, its time evolution operator is generally non-unitary, thereby breaking probability conservation and the standard geometric structure of the state space. Pseudo-Hermitian quantum mechanics provides a systematic resolution to this problem. Its basic idea is not to abandon the Hermiticity principle of quantum mechanics, but rather to treat the inner product structure itself as part of the theory, reconstructing the physical Hilbert space by an appropriate choice of a physical metric.

Specifically, if there exists a linear, Hermitian, and invertible operator $\eta$ such that the Hamiltonian satisfies
\begin{align}
H^\dagger = \eta H \eta^{-1},
\end{align}
then $H$ is called a pseudo-Hermitian operator~\cite{doi:10.1142/S0219887810004816}. This condition indicates that $H$ and its Hermitian conjugate are equivalent under a similarity transformation, imposing strict constraints on its spectral structure. However, in general, $\eta$ does not guarantee the positivity of the inner product. Therefore, pseudo-Hermiticity alone is insufficient to establish a complete probabilistic framework for quantum mechanics.

When the metric operator is further required to be positive-definite and Hermitian, denoted as $\eta_{+}$, the system enters the quasi-Hermitian regime. In this case, a new physical inner product can be defined via $\eta_{+}$:
\begin{align}
(\phi,\psi)_{\eta_{+}} := \langle \phi | \eta_{+} | \psi \rangle,
\end{align}
thereby introducing a new Hilbert space structure $(\mathcal H,(\cdot,\cdot)_{\eta_{+}})$ on the same vector space. Under this physical inner product, quasi-Hermitian systems exhibit real energy spectra and possess a complete and consistent probabilistic interpretation for their dynamics~\cite{Das_2011}.

In the quasi-Hermitian framework, the concepts of ``self-adjointness" and ``observables" must be generalized accordingly. Defining the $\eta_{+}$-adjoint operation
\begin{align}
A^{\ddagger} := \eta_{+}^{-1} A^\dagger \eta_{+},
\end{align}
physical observables are required to satisfy the $\eta_{+}$-self-adjoint condition $O^{\ddagger}=O$. This definition ensures the reality of expectation values for observables and the consistency of the probabilistic interpretation under the physical inner product. Thus, quasi-Hermitian quantum mechanics does not abandon the fundamental principle of Hermiticity; rather, it extends it from the fixed Dirac inner product to the physical inner product determined by the metric operator.

It is worth noting that in many practical problems, the Hamiltonian depends on a set of external parameters $\lambda$, and the compatible positive-definite metric operator may also vary with these parameters, i.e., $\eta_+ = \eta_+(\lambda)$.
This parameter dependence has a direct geometric meaning. Physically, the metric operator $\eta_+(\lambda)$ specifies the rule by which lengths, angles, and overlaps are measured in the physical Hilbert space. As $\eta_+(\lambda)$ changes along a parameter path, the underlying vector space may remain the same, but the physical notion of orthogonality and overlap changes with $\lambda$. This is the basic geometric ingredient that distinguishes the quasi-Hermitian construction from the Hermitian one.

In such cases, although the system can be considered quasi-Hermitian at each fixed parameter point, the inner product structure of the physical Hilbert space itself varies along the parameter path. The primary goal of this paper is to systematically investigate the Hilbert-space geometry introduced by this parameter-dependent inner product structure and to explore its influence on the geometric phase of mixed states.

\subsection{Spectral Structure and Equilibrium Density Matrix of Quasi-Hermitian Systems}

It must be emphasized that the self-adjointness of $H$ established above is with respect to the $\eta_{+}$-inner product; under the standard Dirac inner product, $H$ generally remains non-Hermitian. Consequently, even with a real spectrum, its eigenstate structure differs fundamentally from that of Hermitian systems.

Specifically, let the right eigenstates of the quasi-Hermitian Hamiltonian $H$ satisfy
\begin{align}
H|\Psi_n\rangle = E_n |\Psi_n\rangle .
\end{align}
Since the spectrum $E_n$ is real, one can correspondingly introduce the left eigenstates of $H^\dagger$:
\begin{align}
H^\dagger |\Phi_n\rangle = E_n |\Phi_n\rangle .
\end{align}
Under the Dirac inner product, these two sets of eigenstates generally do not coincide. However, in the non-degenerate case, they can be chosen to satisfy the biorthogonality relations
\begin{align}
\langle \Phi_m | \Psi_n \rangle = \delta_{mn}, \qquad
\sum_n |\Psi_n\rangle \langle \Phi_n| = \mathbb{I}.
\end{align}
This biorthogonal representation, constituted by left and right eigenstates, is a standard structure in quasi-Hermitian spectral theory and provides a natural framework for achieving a consistent spectral decomposition in the quasi-Hermitian representation.

In quasi-Hermitian systems, there exists a direct connection between the biorthogonal structure and the physical inner product. Since $H$ satisfies the self-adjoint condition $H^{\ddagger}=H$ under the $\eta_{+}$-inner product, the left and right eigenstates can be related through the metric operator. Under an appropriate gauge choice, we may take
\begin{align}
|\Phi_n\rangle = \eta_{+} |\Psi_n\rangle ,
\end{align}
so that the right eigenstates satisfy the orthonormality relation under the physical inner product:
\begin{align}
(\Psi_m,\Psi_n)_{\eta_{+}} = \langle \Psi_m | \eta_{+} | \Psi_n \rangle = \delta_{mn}.
\end{align}
This demonstrates that the biorthogonal representation is not in conflict with the physical Hilbert space; rather, it is a concrete manifestation of the positive-definite inner product structure within the non-Hermitian representation of quasi-Hermitian systems.

Under the biorthogonal spectral structure described above, the quasi-Hermitian Hamiltonian can be decomposed as
\begin{align}\label{HEB}
H = \sum_n E_n \, |\Psi_n\rangle \langle \Phi_n| ,
\end{align}
a form analogous to the spectral decomposition in Hermitian systems, but with orthogonality in the state space characterized by the biorthogonal relations.

We now turn to the statistical description of mixed states within the quasi-Hermitian framework. In direct analogy with the Hermitian case, the equilibrium density matrix is defined as
\begin{align}
\rho \equiv \frac{\me^{-\beta H}}{Z},\qquad
Z = \sum_n \me^{-\beta E_n} = \mathrm{Tr}_{\eta_{+}}\!\left(\me^{-\beta H}\right),
\end{align}
where $\beta = 1/T$ is the inverse temperature. Here the physical trace $\mathrm{Tr}_{\eta_+}$ is defined by
\begin{align}
\mathrm{Tr}_{\eta_{+}}(A) :
= \sum_n \langle \Psi_n | \eta_{+} A | \Psi_n \rangle
= \sum_n \langle \Phi_n | A | \Psi_n \rangle,
\end{align}
which, in the spectral representation, coincides with the biorthogonal trace~\cite{doi:10.1142/S0219887810004816}. Using the quasi-Hermitian condition $H^{\ddagger}=H$, one finds that the density matrix satisfies $\rho^{\ddagger} = \rho$, confirming that $\rho$ is self-adjoint under the physical inner product and therefore represents a legitimate physical mixed state.

Furthermore, as an analytic function of $H$, the density matrix admits a spectral decomposition in the biorthogonal basis analogous to Eq.~\eqref{HEB}:
\begin{align}\label{DMEB}
\rho = \sum_n p_n \, |\Psi_n\rangle \langle \Phi_n| , \qquad
p_n = \frac{\me^{-\beta E_n}}{Z},
\end{align}
where the weights $p_n$ are identical to the Gibbs distribution in Hermitian systems. The eigenvalues are non-negative and sum to unity, ensuring the standard probabilistic interpretation.

Thus, we have established the basic statistical description of mixed states within the quasi-Hermitian framework. Although the Hamiltonian is non-Hermitian in the Dirac representation, the introduction of a positive-definite metric operator restores self-adjointness and probabilistic consistency in the physical Hilbert space, yielding a spectral representation built upon the biorthogonal basis. This structure provides the necessary foundation for the subsequent introduction of the geometric phase of mixed states.

\section{Theoretical Framework of Uhlmann Phase in Quasi-Hermitian Systems}

When a quasi-Hermitian system depends on a set of tunable external parameters $\lambda$, both the Hamiltonian and the metric operator typically become functions of these parameters, i.e., $H(\lambda)$ and $\eta_{+}(\lambda)$. Together, they determine the physical inner product structure along the parameter path. Unlike in Hermitian systems, in the quasi-Hermitian case the inner product of the physical Hilbert space itself evolves with the parameters. This parameter-dependent inner product fundamentally alters the geometric properties of the state space, raising a natural question: how should one define and characterize the geometric phase acquired by a mixed state when such a system undergoes a cyclic evolution in parameter space?

To address this question, we develop a geometric framework for mixed states in quasi-Hermitian systems by generalizing Uhlmann's construction of geometric phases for density matrices. The central idea is to represent each mixed state $\rho$ by an amplitude operator $W$ satisfying $\rho = WW^{\ddagger}$, where the $\ddagger$-operation is defined with respect to the parameter-dependent metric $\eta_+(\lambda)$. This representation lifts the evolution of mixed states in the parameter space to a path in a higher-dimensional amplitude space.

The price we pay for this lift is the introduction of a gauge redundancy: different amplitudes related by $W \to WU$ with $U^{\ddagger}U = \mathbb I$ represent the same physical mixed state. This gauge freedom, governed by the quasi-unitary group $U_{\eta_+}(N)$, provides the language needed to separate geometric effects from dynamical ones.

By imposing a parallel transport condition that selects a ``natural" lift of the path through minimizing the distance between neighboring amplitudes, we obtain a connection whose holonomy along a closed loop defines the geometric phase, the quasi-Hermitian Uhlmann phase. The key distinction from the Hermitian case is that the gauge group itself depends on the parameters through the metric $\eta_+(\lambda)$, introducing new geometric features that we will explore in detail.
\subsection{Purification and Gauge Structure of Quasi-Hermitian Mixed States}
\label{subsec:qh_purification}

To make this construction explicit, we introduce the amplitude operator $W$ (also called a purification) satisfying $\rho = W W^{\ddagger}$, with $\ddagger$ the $\eta_+$-adjoint operation defined above. This representation thus lifts the density matrix $\rho$ to a point $W$ in the purification space.

The purification representation $\rho = W W^{\ddagger}$ is not unique. If $W$ is a purification of $\rho$, then for any operator $U$ satisfying the quasi-unitary condition $U^{\ddagger} U = \mathbb{I}$, the product $WU$ describes the same mixed state, corresponding to the same density matrix $\rho$. Thus, we have the equivalence relation
\begin{align}
W \sim WU .
\label{eq:gauge_equiv_W}
\end{align}

The gauge group preserving the physical inner product is the quasi-unitary group $U_{\eta_{+}}(N)$, which replaces the unitary group $U(N)$ of the Hermitian case. For a full-rank density matrix, any purification can be written as
\begin{align}
W = \sqrt{\rho} U
\label{eq:W_sqrt_rho}
\end{align}
where $\sqrt{\rho}$ is the positive square root of $\rho$ in the quasi-Hermitian Hilbert space. Under biorthogonal spectral decomposition, $\sqrt{\rho}=\sum_n \sqrt{p_n}\,|\Psi_n\rangle\langle\Phi_n|$, encoding the spectral structure and statistical weights. The factor $U \in U_{\eta_{+}}(N)$ characterizes the gauge degrees of freedom, playing a role analogous to the phase factor in pure-state geometry.

To complete the geometric description, we note that, for each fixed parameter value $\lambda$, the corresponding
purification space carries a well-defined Hilbert-space structure equipped with the $\eta_+(\lambda)$-dependent Hilbert-Schmidt inner product. Since the metric operator $\eta_+(\lambda)$ generally varies along the parameter manifold, these purification spaces should more precisely be regarded as forming a parameter-dependent family of Hilbert spaces $\mathcal{H}^{\eta_{+}}_{W}(\lambda)$ rather than a single Hilbert space equipped with a fixed global inner product.

A suitable inner product on the corresponding purification space, namely the quasi-Hermitian Hilbert-Schmidt inner product, can be defined in direct analogy with the standard Hermitian case~\cite{ChruscinskiJamiolkowski2004} as
\begin{align}
(W_1, W_2)_{\mathrm{HS}}^{\eta_{+}}
:=
\mathrm{Tr}_{\eta_{+}}\!\left( W_1^{\ddagger} W_2 \right).
\label{eq:qh_HS_product}
\end{align}
One can verify that, for each fixed parameter value $\lambda$, this defines a Hermitian scalar product fully compatible with the underlying quasi-Hermitian inner-product structure induced by the metric operator
$\eta_+(\lambda)$.

In the Hermitian case, the purification space naturally forms a principal fiber bundle with fixed structure group $U(N)$.
In the quasi-Hermitian setting, however, the situation becomes more subtle due to the parameter dependence of the metric operator $\eta_+(\lambda)$.

Rather than organizing the purification space into a principal bundle with a fixed structure group, it is more appropriate to regard the present construction as defining a parameter-dependent family of fibers associated with the equivalence relation $W \sim WU$.

The natural projection is
\begin{align}
\pi:\ S_N\to D_N^{\;N},\qquad
\pi(W):=WW^{\ddagger}=\rho .
\label{eq:bundle_projection}
\end{align}

At any point $\rho$, the fiber consists of all equivalent purifications,
\begin{align}
\pi^{-1}(\rho)
=
\left\{\,W=\sqrt{\rho}\,U\ \big|\
U\in U_{\eta_{+}}(N)\,\right\}.
\label{eq:fiber}
\end{align}

At this point, it is important to clarify the mathematical nature of the structure group in this construction.
While the fiber at a point $\rho$ is given by $W = \sqrt{\rho}\,U$ with $U$ satisfying $U^{\ddagger}U = \mathbb I$, the operator $\ddagger$ depends explicitly on the metric $\eta_+(\lambda)$.
Consequently, the set of such $U$ forms the quasi-unitary group $U_{\eta_+(\lambda)}(N)$, which itself varies with the parameter $\lambda$.

Strictly speaking, a principal bundle is defined with a fixed structure group $G$ that does not depend on the base point.
Therefore, our construction is not a principal bundle in the conventional sense.
Instead, it can be more rigorously understood as a reduction of structure group of a larger, fixed principal bundle, specifically, a $GL(N, \mathbb{C})$-bundle, where the reduction to the subgroup $U_{\eta_+(\lambda)}(N)$ is parameter-dependent.
The variation of $\eta_+(\lambda)$ along the parameter space means that we are effectively moving through different reductions of the same ambient $GL(N, \mathbb{C})$-bundle.

For the purpose of constructing the Uhlmann phase via parallel transport along a given path $\rho(\lambda)$, this subtlety does not affect our local computations.
At each fixed $\lambda$, $U_{\eta_+(\lambda)}(N)$ serves as the effective symmetry group governing the fiber directions.
The parallel transport condition and the resulting connection to be introduced below are defined with respect to this $\lambda$-dependent group, which is sufficient for defining the pathwise holonomy and the geometric phase along a closed loop in the extended parameter space.

This construction provides a path-dependent geometric framework for describing the transport of purification amplitudes in quasi-Hermitian systems: the base space is the manifold of density matrices, the total space is the amplitude space, and the fibers correspond to parameter-dependent quasi-unitary gauge degrees of freedom.

Within this framework, the holonomy should be understood in a pathwise sense. Once an initial purification is specified, the quasi-Hermitian parallel-transport condition selects a horizontal lift along the loop, and the resulting holonomy is defined by the endpoint mismatch after one cycle. Thus, it is not the holonomy of a globally defined principal connection with a fixed structure group, but a pathwise holonomy associated with the metric-dependent family of fibers along the loop. This pathwise structure is sufficient for defining the Uhlmann phase along a closed parameter loop, and no global bundle description over the entire parameter space is required. We now turn to the parallel-transport condition that determines the corresponding pathwise connection.

\subsection{Uhlmann Phase in Quasi-Hermitian Systems}
\label{subsec:qh_uhlmann_connection}

To define the geometric phase, we select a distinguished pathwise lift of
$\rho(\lambda)$ to the amplitude space by imposing an appropriate parallel-transport condition within the path-dependent quasi-unitary gauge structure.
This is achieved by imposing a parallel transport condition that minimizes the Hilbert-Schmidt distance between neighboring purifications~\cite{ourPRB20}.

To this end, we introduce the Hilbert-Schmidt distance in the quasi-Hermitian purification space. For two purification operators $W_1$ and $W_2$, we define
\begin{align}
D_{\mathrm{HS}}^2(W_1,W_2)
:=
\mathrm{Tr}_{\eta_{+}}\!\left[
(W_1-W_2)^{\ddagger}(W_1-W_2)
\right].
\end{align}
This metric-weighted distance is the natural Hilbert--Schmidt distance associated with the physical inner product defined by $\eta_+(\lambda)$. Its extremization leads to the quasi-Hermitian parallel-transport condition. For a fixed metric, the construction reduces to the standard Uhlmann construction in the corresponding physical Hilbert space, whereas for a parameter-dependent metric the adjoint operation $\ddagger$ must be understood as a $\lambda$-dependent operation determined by the local metric $\eta_+(\lambda)$. The overlap between the initial purification and the parallel-transported final purification then gives the quasi-Hermitian Uhlmann amplitude.

Since purifications possess gauge redundancy, their lifted paths can be adjusted by $U(\lambda)$. This condition selects, within the gauge degrees of freedom, the lift that minimizes this distance between neighboring purifications. Its continuous limit equivalently yields
\begin{align}
W^{\ddagger}(\lambda)\,\dot W(\lambda)
=
\dot{(W^{\ddagger})}(\lambda)\,W(\lambda),
\quad
\dot{(\,)}\equiv\frac{\dif}{\dif\lambda}.
\label{eq:qh_uhlmann_parallel}
\end{align}

A detailed derivation can be found in Appendix~\ref{app:derivation_parallel_condition}. This condition is a natural generalization of the standard Uhlmann parallel condition to the $\eta_{+}$-inner product structure. From a geometric perspective, Eq. \eqref{eq:qh_uhlmann_parallel} selects a horizontal curve in the purification space, thereby distinguishing horizontal variations from gauge (vertical) variations along the fiber directions.

Here the parameter $\lambda$ labels the geometric path along which both the density matrix and the metric are transported. Along this path, the adjoint operation and the Hilbert--Schmidt structure are evaluated with the local metric $\eta_+(\lambda)$. If the external parameters are driven in real time, $\lambda=\lambda(t)$, the time-evolution generator must satisfy the norm-conservation condition imposed by the time-dependent metric. For a state obeying $\mi\partial_t|\psi(t)\rangle=K(t)|\psi(t)\rangle$, conservation of the physical norm $\langle\psi(t)|\eta_+(\lambda(t))|\psi(t)\rangle$ gives
\begin{align}
K^\dagger(t)\eta_+(\lambda(t))-\eta_+(\lambda(t))K(t) = \mi\,\frac{\dif}{\dif t}\eta_+(\lambda(t)).
\end{align}
Thus, when $\frac{\dif}{\dif t}\eta_+(\lambda(t))\neq0$, the quasi-Hermiticity condition $H^\dagger(t)\eta_+(\lambda(t))=\eta_+(\lambda(t))H(t)$ cancels the contribution from the state evolution, but it does not cancel the contribution from the explicit time variation of the metric. Norm-preserving time evolution then requires the stronger condition displayed above. This issue pertains to real-time dynamical realizations. By contrast, the quasi-Hermitian Uhlmann construction is formulated as a geometric parallel transport in parameter space, for which no time-evolution generator needs to be specified.

Along a parallel transport path, the quasi-Hermitian Uhlmann connection is given by
\begin{align}
\mathcal{A}_{\text{U}}^{\eta_{+}}(\lambda)=-\dif U(\lambda)\,U^{\ddagger}(\lambda).
\label{eq:qh_A_def}
\end{align}
This expression takes a form directly analogous to its Hermitian counterpart~\cite{ourPRB20}, with the crucial distinction that the adjoint operation and the resulting connection are now defined with respect to the physical metric $\eta_+$ at each point along the parameter path.

For a parameter-independent metric, the quasi-unitary group $U_{\eta_+}(N)$ remains fixed along the path, and Eq.~\eqref{eq:qh_A_def} satisfies $\left(\mathcal A_{\text{U}}^{\eta_+}\right)^{\ddagger}=-\mathcal A_{\text{U}}^{\eta_+}$, so that the connection takes values in its Lie algebra and is fully compatible with the quasi-Hermitian inner product structure.
When the metric depends on the parameters, $\eta_+(\lambda)$ and the corresponding quasi-unitary group $U_{\eta_+(\lambda)}(N)$ vary along the path. Since the $\ddagger$ operation itself also depends on $\lambda$, $\dif(U^{\ddagger})$ generally differs from $(\dif U)^{\ddagger}$. Therefore, differentiating $U^{\ddagger}U=I$ gives $\dif(U^{\ddagger})U+U^{\ddagger}\dif U=0$, rather than an anti-self-adjointness condition for $\mathcal A_{\text{U}}^{\eta_+}$ in a fixed Lie algebra. Accordingly, $\mathcal A_{\text{U}}^{\eta_+}$ should be more properly understood as a pathwise connection compatible with the parameter-dependent family of quasi-unitary groups $U_{\eta_+(\lambda)}(N)$.
 Equivalently, the above expression can be rewritten as the pathwise parallel evolution equation for the phase factor:
\begin{align}
\dif U + \mathcal{A}_{\text{U}}^{\eta_{+}}U = 0.
\label{eq:qh_covariant_parallel_U}
\end{align}

Integrating this equation along a loop $\mathcal{C}:=\{\rho(\lambda) \mid 0 \le \lambda \le \tau, \, \rho(0)=\rho(\tau)\}$ with path ordering yields
\begin{align}
U(\tau)
=
\mathcal P
\exp\!\left(
-\int_{0}^{\tau} \mathcal{A}_{\text{U}}^{\eta_{+}}(\lambda)
\right)
U(0),
\label{eq:qh_holonomy_563}
\end{align}
where $\mathcal P$ denotes the path-ordering operator. This expression gives the Uhlmann holonomy obtained along a closed parameter path.

Geometrically, $\mathcal A_{\text{U}}^{\eta_+}(\lambda)$ should be understood as a path-dependent connection one-form defined along the purification lift of the density-matrix trajectory $\rho(\lambda)$, rather than as the pullback of an Ehresmann connection on a globally defined principal bundle with fixed structure group \cite{ourPRB20}.
The holonomy associated with this pathwise connection is gauge covariant, while the resulting Uhlmann amplitude and Uhlmann phase are gauge invariant. They reflect the intrinsic geometry of the mixed-state path together with the metric-dependent overlap structure.

The explicit form of the quasi-Hermitian Uhlmann connection can also be derived from the parallel transport condition. Substituting the gauge decomposition form of the purification into the parallel evolution condition \eqref{eq:qh_uhlmann_parallel} and utilizing the basic properties of the quasi-Hermitian adjoint operation, we obtain a deterministic equation satisfied by $\mathcal{A}_{\text{U}}^{\eta_{+}}$:
\begin{align}
\rho\,\mathcal{A}_{\text{U}}^{\eta_{+}}+\mathcal{A}_{\text{U}}^{\eta_{+}}\rho
=
-\bigl[\dif\sqrt{\rho},\sqrt{\rho}\bigr].
\label{eq:qh_sylvester_final}
\end{align}
This equation demonstrates that the quasi-Hermitian Uhlmann connection is completely determined by the variation of the base space path $\rho(\lambda)$ and is independent of specific gauge choices. A detailed derivation is provided in Appendix~\ref{app:derivation_uhlmann_connection}.

Substituting the spectral decomposition of the density matrix, Eq.~\eqref{DMEB}, and working in the biorthogonal eigenbasis, the above equation can be solved element-wise, yielding the explicit expression for the quasi-Hermitian Uhlmann connection:
\begin{align}
\mathcal{A}_{\text{U}}^{\eta_{+}}
=
-\sum_{i,j}
|\Psi_i\rangle
\frac{\langle \Phi_i|\,[\dif\sqrt{\rho},\sqrt{\rho}]\,|\Psi_j\rangle}{p_i+p_j}
\langle \Phi_j|.
\label{eq:qh_A_explicit_final}
\end{align}

With this expression and Eq.~(\ref{eq:qh_holonomy_563}), we obtain the Hilbert-Schmidt inner product between the initial and final purifications,
\begin{align}\label{QHUPG}
\mathcal{G}
:=
\bigl( W(0), W(\tau) \bigr)_{\mathrm{HS}}^{\eta_{+}}
=
\mathrm{Tr}_{\eta_{+}}
\!\left(
\rho(0)\,
U(\tau)
\right),
\end{align}
a quantity known as the Uhlmann amplitude, which directly yields the quasi-Hermitian Uhlmann phase
\begin{align}\label{QHUP}
\theta_{\text{U}}^{\eta_{+}}
=\arg \mathcal{G}=
\arg\,
\mathrm{Tr}_{\eta_{+}}
\!\left[
\rho(0)\,
U(\tau)
\right].
\end{align}
When $\eta_+$ reduces to the identity, the above results reduce to the Uhlmann phase in Hermitian systems.

\begin{figure}[th]
\centering
\includegraphics[width=3.3in]{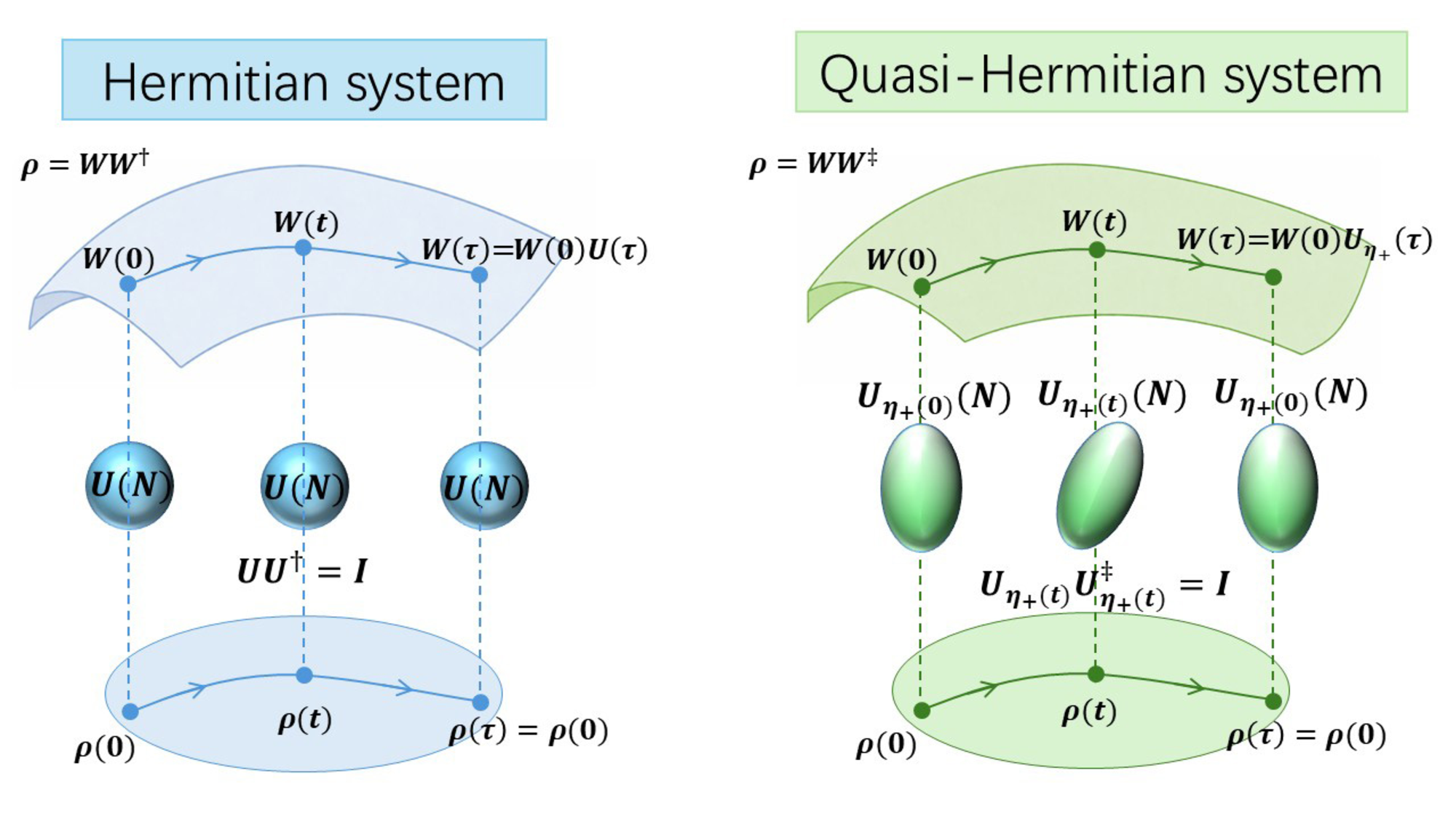}
\caption{Schematic comparison of the Hermitian and quasi-Hermitian constructions of the Uhlmann process along a closed loop $\rho(0)\rightarrow\rho(\tau)=\rho(0)$.}
\label{Fig00}
\end{figure}

Geometrically, this construction can be illustrated by the schematic comparison shown in Fig.~\ref{Fig00}.
Along a closed trajectory $\rho(0)\rightarrow\rho(\tau)=\rho(0)$ in the space of density matrices, the corresponding lifted curve $W(\lambda)$ represents the Uhlmann parallel transport of purification amplitudes in the enlarged space.
In the Hermitian case, the associated fibers are characterized by a fixed structure group $U(N)$, which is represented schematically by spheres.
By contrast, in the quasi-Hermitian case the gauge sectors are described by the parameter-dependent quasi-unitary group $U_{\eta_{+}(\lambda)}(N)$ due to the variation of the metric operator $\eta_{+}(\lambda)$, illustrated by ellipsoids with different shapes along the parameter path.

Accordingly, the lifted amplitudes are compared with a local inner product that changes along the path, rather than with a single fixed Hilbert--Schmidt structure.
After one cyclic evolution, the endpoint mismatch between $W(0)$ and $W(\tau)$ therefore contains information not only about the loop $\rho(\lambda)$, but also about the metric-dependent overlap structure of the quasi-Hermitian Hilbert space. It encodes the pathwise Uhlmann holonomy and gives rise to the geometric phase defined in Eq.~(\ref{QHUP}).

To analyze the zero structure of $\mathcal{G}$ and its possible nonanalytic behavior, we introduce, by analogy with the thermodynamic free-energy density, the geometrical generating function~\cite{OurPRB20b}
\begin{align}
g
=
-
\lim_{L \to \infty}
\frac{1}{L}
\ln
\left|
\mathcal{G}
\right|^{2},
\end{align}
where $L$ denotes the system size. Nonanalyticities of $g$ occur precisely when $\mathcal{G}$ approaches zero.

Finally, it is instructive to examine the topological content of the quasi-Hermitian Uhlmann connection.
From Eq.~(\ref{eq:qh_A_explicit_final}), the Uhlmann connection $\mathcal{A}_{\text{U}}^{\eta_+}$ is constructed from the commutator $[\dif\sqrt{\rho},\sqrt{\rho}]$.
Accordingly, in the biorthogonal eigenbasis at each point along the parameter path, $\mathcal{A}_{\text{U}}^{\eta_+}$ has vanishing diagonal elements and thus satisfies
\begin{align}
\mathrm{Tr}_{\eta_+}\mathcal{A}_{\text{U}}^{\eta_+}=0 .
\end{align}

Formally, a curvature-like two-form associated with the induced connection can be introduced,
\begin{align}
\mathcal{F}_{\text{U}}^{\eta_+}
=
\dif \mathcal{A}_{\text{U}}^{\eta_+}
+
\mathcal{A}_{\text{U}}^{\eta_+}
\wedge
\mathcal{A}_{\text{U}}^{\eta_+},
\end{align}
which inherits this property and remains traceless,
\begin{align}
\mathrm{Tr}_{\eta_+}\,\mathcal{F}_{\text{U}}^{\eta_+} = 0.
\end{align}
Accordingly, a first-Chern-type quantity constructed from this curvature-like two-form satisfies
\begin{align}
c_1^{(\mathrm{type})}
\propto
\int \mathrm{Tr}_{\eta_+}\,\mathcal{F}_{\text{U}}^{\eta_+}
= 0,
\end{align}
indicating that no nontrivial first-Chern-type invariant arises from the present path-dependent connection structure.

In particular, despite the non-Hermitian nature of the Hamiltonian and the presence of a parameter-dependent metric $\eta_+(\lambda)$, the quasi-Hermitian extension does not generate a nontrivial topological invariant at the level of this curvature-like construction.

Instead, the nontrivial features uncovered in this work originate from the geometric structure of the connection itself, especially its dependence on the metric and thermal mixing, rather than from a change in topological class in the sense of characteristic classes of a globally defined principal bundle.

It is important to emphasize that the present construction is formulated along parameterized paths in the space of density matrices. Accordingly, the associated connection and holonomy should be understood as path-dependent geometric structures rather than global bundle-level topological invariants.

\section{Two-Level Models: Parameter-Independent and Parameter-Dependent Metrics}
We now visualize the abstract formalism developed above with two analytically solvable two-level models. These examples are chosen to highlight the contrasting roles of the metric: in the first model, the metric is parameter-independent, whereas in the second it varies with external parameters. As we shall see, this distinction has profound consequences for the mixed-state geometric phase.
\subsection{A Simple Two-Level Model with Parameter-Independent Metric}
\label{sec:two_level_model}
\subsubsection{Model Definition}
We consider a two-level quasi-Hermitian Hamiltonian parameterized by an angle $\theta$ and a real deformation parameter $t>0$:
\begin{align}
H(\theta) = \begin{pmatrix}
\cos \theta & t \sin \theta \\
\frac{\sin \theta}{t} & -\cos \theta
\end{pmatrix}. \label{eq:two_level_h}
\end{align}
The eigenvalues of $H$ are given by $E_\pm = \pm 1$,
independent of $\theta$, indicating that the Hamiltonian describes an isospectral family. The corresponding right eigenstates are
\begin{align}
|\psi_+\rangle = \begin{pmatrix} \cos \frac{\theta}{2} \\ \frac{\sin \frac{\theta}{2}}{t} \end{pmatrix}, \quad
|\psi_-\rangle = \begin{pmatrix} \sin \frac{\theta}{2} \\ -\frac{\cos \frac{\theta}{2}}{t} \end{pmatrix},
\end{align}
and the left eigenstates take the form
\begin{align}
\langle \phi_+| = \begin{pmatrix} \cos \frac{\theta}{2} & t \sin \frac{\theta}{2} \end{pmatrix}, \quad
\langle \phi_-| = \begin{pmatrix} \sin \frac{\theta}{2} & -t \cos \frac{\theta}{2} \end{pmatrix}.
\end{align}
These eigenstates satisfy the biorthogonality condition $\langle \phi_m | \psi_n \rangle = \delta_{mn}$. To ensure quasi-Hermiticity, we introduce the positive-definite metric operator
\begin{align}
\eta_{+} = \begin{pmatrix}
1 & 0 \\
0 & t^2
\end{pmatrix}, \label{eq:eta_two_level}
\end{align}
which satisfies $H^\dagger = \eta_{+} H \eta_{+}^{-1}$. The physical inner product is thus defined by $(\cdot,\cdot)_{\eta_{+}} = \langle \cdot | \eta_{+} | \cdot \rangle$.
\subsubsection{Uhlmann Connection and Phase}
We now examine the finite-temperature Uhlmann phase associated with this model. The system is assumed to be in thermal equilibrium at inverse temperature $\beta = 1/T$, with the Gibbs state $\rho(\theta) = \me^{-\beta H(\theta)}/Z$. At the initial parameter value $\theta = 0$, the Hamiltonian is diagonal, and the density matrix simplifies to
\begin{align}
\rho(0) = \frac{1}{\me^{\beta} + \me^{-\beta}} \begin{pmatrix} \me^{-\beta} & 0 \\ 0 & \me^{\beta} \end{pmatrix},
\end{align}
where we adopt the convention that the eigenvalue $+1$ corresponds to the higher energy state.
Following the quasi-Hermitian Uhlmann formalism developed in Sec.~III, we compute the Uhlmann connection $\mathcal{A}_{\text{U}}^{\eta_{+}}$ along a path in $\theta$. For the present model, a straightforward calculation using the general expression \eqref{eq:qh_A_explicit_final} yields
\begin{align}
\mathcal{A}^{\eta_{+}}_{\text{U}} = x \begin{pmatrix} 0 & -\frac{t}{2} \\ \frac{1}{2t} & 0 \end{pmatrix} \dif\theta, \qquad
x = \frac{2 - \me^{\beta} - \me^{-\beta}}{\me^{\beta} + \me^{-\beta}}.
\label{eq:uhlmann_connection_two_level}
\end{align}

Consider a closed path in parameter space that winds $\Omega$ times around the circle $\theta \in [0,2\pi\Omega]$. The corresponding Uhlmann holonomy is obtained by path-ordering the exponential of the integrated connection:
\begin{align}
U(2\pi\Omega) &= \exp\!\left( -2\pi\Omega x \begin{pmatrix} 0 & -t/2 \\ 1/(2t) & 0 \end{pmatrix} \right)\notag\\
&=\begin{pmatrix}
\cos(\pi\Omega x) & t \sin(\pi\Omega x) \\[2pt]
\displaystyle -\frac{\sin(\pi\Omega x)}{t} & \cos(\pi\Omega x)
\end{pmatrix}.
\end{align}
Using $x = -(1-1/\cosh\beta)$, the Uhlmann amplitude and Uhlmann phase are respectively given by
\begin{align}
\mathcal{G} = \cos\!\left( \pi\Omega \left(1 - \frac{1}{\cosh\beta}\right) \right),
\end{align}
and
\begin{align}
\theta_{\text{U}}^{\eta_{+}}
= \arg\!\left[ \cos\!\left( \pi\Omega \left(1 - \frac{1}{\cosh\beta}\right) \right) \right].
\label{eq:uhlmann_phase_two_level}
\end{align}
Since the cosine function is real, $\theta_{\text{U}}^{\eta_{+}}$ is either $0$ or $\pi$ (mod $2\pi$), depending on the sign of $\mathcal{G}$.
\subsubsection{Finite-Temperature Geometric Phase Transitions}
\label{sec:topological_transitions}
The Uhlmann phase $\theta_{\text{U}}^{\eta_{+}}$ encodes the geometric properties of thermal states, exhibiting abrupt jumps at critical temperatures that signal geometric phase transitions. These discontinuities occur when the cosine argument in Eq.~\eqref{eq:uhlmann_phase_two_level} satisfies
\begin{align}
\pi\Omega \left(1 - \frac{1}{\cosh\beta_{n,c}}\right) = \left(n+\tfrac{1}{2}\right)\pi, \quad n \in \mathbb{Z}.
\end{align}
At these critical temperatures $T_{n,c}=1/\beta_{n,c}$, the Uhlmann amplitude $\mathcal{G}$ vanishes, rendering $\theta_{\text{U}}^{\eta_{+}}$ ill-defined and marking the boundary between two distinct Uhlmann-phase sectors ($\theta_{\text{U}}^{\eta_{+}}=0$ and $\theta_{\text{U}}^{\eta_{+}}=\pi$).
For a single winding ($\Omega = 1$), the condition reduces to
\begin{align}
\frac{1}{\cosh\beta_{n,c}} = \tfrac{1}{2} - n, \quad n \in \mathbb{Z}.
\end{align}
Since $1/\cosh\beta_{n,c}$ is restricted to the interval $(0,1]$, only the $n=0$ branch is admissible, giving $\beta_c=\operatorname{arcosh}2$. Across the critical point, the Uhlmann phase changes from the low-temperature nontrivial value $\theta_{\text{U}}^{\eta_+}=\pi$ to the high-temperature trivial value $\theta_{\text{U}}^{\eta_+}=0$.

\begin{figure}[th]
\centering
\includegraphics[width=3.3in]{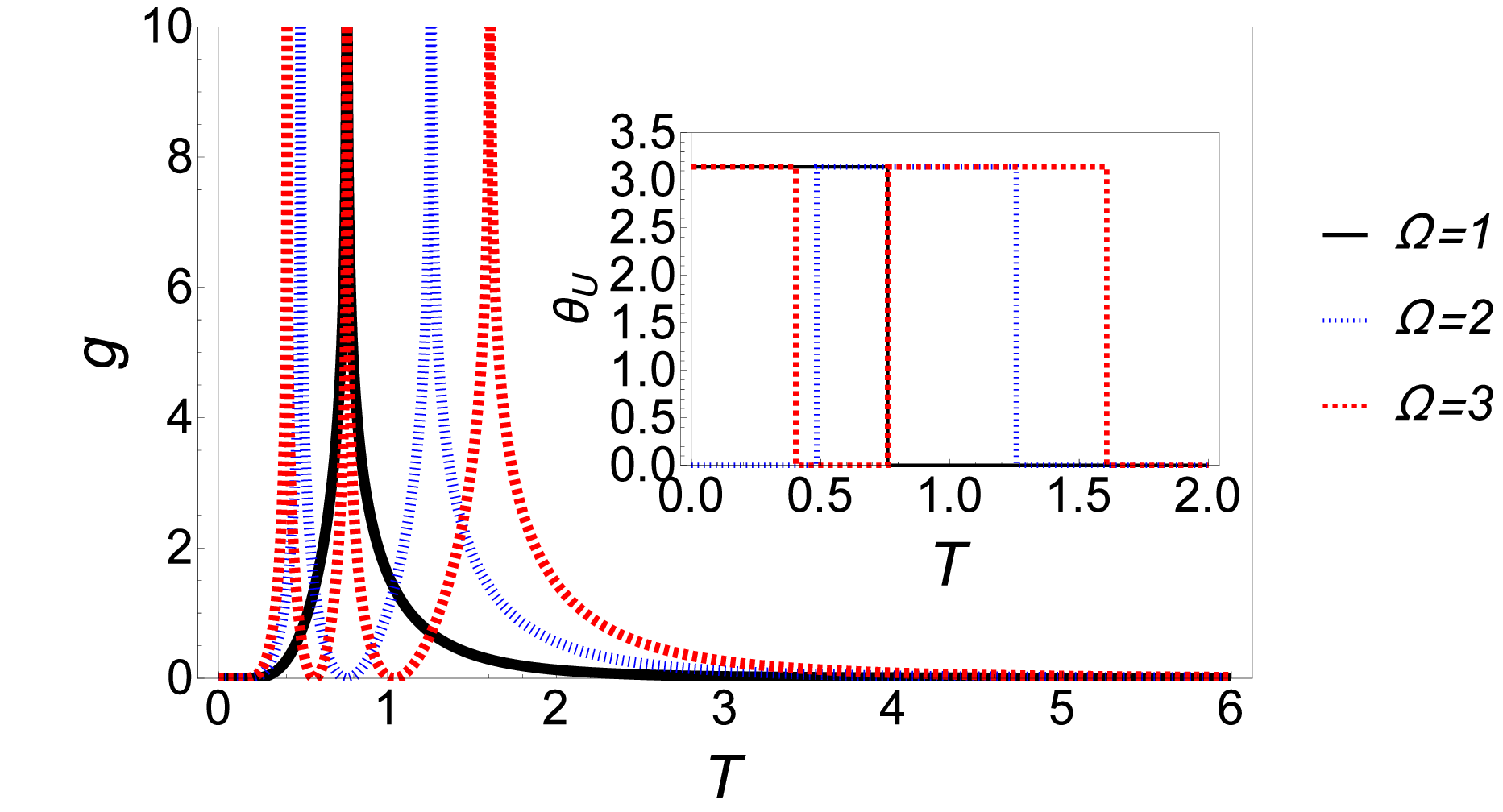}
\caption{The Uhlmann phase $\theta_{\text{U}}^{\eta}$ (inset) and the geometrical generating function $g$ as functions of temperature for $\Omega = 1$ (black solid), $\Omega = 2$ (blue dotted), and $\Omega = 3$ (red dashed). The diverging peaks indicate critical temperatures where the phase jumps by $\pi$.}
\label{Fig0}
\end{figure}

More generally, the winding number $\Omega$ dictates the number of phase transitions. As illustrated in Fig.~\ref{Fig0}, the $\Omega=2$ case exhibits a richer structure: starting from the trivial phase at zero temperature, the system enters a nontrivial phase at an intermediate critical temperature before returning to the trivial phase at a higher temperature. For $\Omega=3$, three successive transitions occur, with nontrivial phases emerging in alternating temperature windows.

These observations reveal a general pattern: the high-temperature limit universally yields a trivial sector of the Uhlmann phase, whereas nontrivial Uhlmann-phase structure survives only within specific finite-temperature windows. The winding number $\Omega$ thus serves as a control parameter governing the complexity of the phase diagram, determining both the number and the locations of geometric-phase transitions across the temperature spectrum.

Accordingly, the above result reduces to that of a Hermitian spin-$\tfrac{1}{2}$ system evolving along a meridional loop~\cite{PhysRevA.104.023303}. This agreement follows from the fact that the metric operator $\eta_+$ in the present model is independent of the external parameters, so that no additional geometric contribution arises from metric variation along the parameter path. Since the inner-product structure remains unchanged along the entire parameter path, quasi-Hermiticity introduces no additional metric-induced geometric contributions, and the mixed-state geometric phase therefore reduces to its
Hermitian counterpart. A more detailed analysis of this equivalence is presented in Appendix~\ref{subsec:qh_vs_h_uhlmann}.

\subsection{A $\mathcal{PT}$-Symmetric Two-Level Model with Parameter-Dependent Metric}
While the previous example demonstrates finite-temperature geometric phase transitions captured by the Uhlmann phase, the quasi-Hermitian coefficient $t$ does not directly influence the value of the Uhlmann phase. We now construct a more intricate example to explicitly showcase the impact of quasi-Hermitian coefficients on the geometric structure of quantum states.
\subsubsection{The $\mathcal{PT}$-Symmetric Two-Level Model and Its Spectral Structure}
\label{sec:PT_model}

We consider a quantum model defined on a two-dimensional Hilbert space. In this setting, any Hermitian $2\times2$ operator can be expressed as a linear combination of Pauli matrices, naturally corresponding to a direction on the Bloch sphere. As a starting point, we introduce a simple reference Hermitian Hamiltonian
\begin{align}
H_{2\times 2}
=
\hat{e}_r\cdot\vec{\sigma}
=
\begin{pmatrix}
\cos\theta & \sin\theta\,\mathrm{e}^{-i\phi} \\
\sin\theta\,\mathrm{e}^{i\phi} & -\cos\theta
\end{pmatrix},
\label{eq:P_2x2}
\end{align}
where $\vec{\sigma}=(\sigma_x,\sigma_y,\sigma_z)$ are the standard Pauli matrices, and the unit vector
\begin{align}
\hat{e}_r
=
(\sin\theta\cos\phi,\;
 \sin\theta\sin\phi,\;
 \cos\theta),
\end{align}
parametrizes the radial direction on the Bloch sphere. The eigenstates of this Hamiltonian correspond to pure states polarized along $\hat{e}_r$, with the spectrum and geometric structure controlled entirely by the spherical angles $(\theta,\phi)$. To avoid confusion with vectors in an arbitrary parameter space $\mathbf{X}$, we use arrow notation $\vec{A}$ exclusively for vectors in three-dimensional real space.

Building on this, a general $\mathcal{PT}$-symmetric $2\times2$ Hamiltonian is characterized by six real parameters ${\bf X}=(\varepsilon,a,b,\theta,\phi,\delta)$ and takes the form
\begin{align}
H_{2\times 2}({\bf X})
=
\varepsilon
+
\left(
a\,\hat{e}_r
+\mi b\cos\delta\,\hat{e}_{\theta}
+\mi b\sin\delta\,\hat{e}_{\varphi}
\right)\cdot\vec{\sigma},
\label{eq:H_PT_2x2}
\end{align}
where $\varepsilon$ represents a global energy shift, while $a$ and $b$ control the coupling strengths along the radial and tangential directions, respectively. The two orthogonal unit vectors
\begin{align}
\hat{e}_{\theta}
&=
(\cos\theta\cos\phi,\;
 \cos\theta\sin\phi,\;
 -\sin\theta),\notag\\
\hat{e}_{\varphi}
&=
(-\sin\phi,\;
 \cos\phi,\;
 0),
\end{align}
span the tangent space of the Bloch sphere at the point $(\theta,\phi)$ \cite{Wang_2010}.
This $\mathcal{PT}$-symmetric Hamiltonian can be understood as a geometric generalization of the Hermitian model $H_{2\times2}=\hat{e}_r\cdot\vec{\sigma}$. The component along the radial direction $\hat{e}_r$ remains real and determines the overall spectral structure, while the tangential components along $\hat{e}_\theta$ and $\hat{e}_\varphi$ are complexified into purely imaginary terms, introducing non-Hermiticity in a controlled manner. This particular complexification ensures that the Hamiltonian retains $\mathcal{PT}$ symmetry while preserving the clear Bloch sphere geometry of the two-dimensional system, thereby providing an ideal minimal model for analyzing mixed-state geometric structures.

The spectrum follows directly from the magnitude of the effective Bloch vector. Working in the orthonormal basis $\{\hat{e}_r,\hat{e}_\theta,\hat{e}_\varphi\}$, the two eigenvalues of $H_{2\times2}({\bf X})$ are found to be
\begin{align}
E_{\pm}
=
\varepsilon \pm \sqrt{a^2-b^2}.
\label{eq:PT_eigenvalues}
\end{align}
Consequently, the system possesses a real, non-degenerate spectrum precisely when $a^2>b^2$, which corresponds to the $\mathcal{PT}$-unbroken phase where quasi-Hermitian quantum mechanics applies. For $a^2<b^2$, the spectrum consists of a complex conjugate pair, and at the critical point $a^2=b^2$, the two eigenvalues coalesce at an exceptional point.

In what follows, we restrict our attention to the quasi-Hermitian regime. Within this parameter region, one can introduce a positive-definite metric operator compatible with the Hamiltonian. In the two-dimensional Hilbert space, its explicit form is
\begin{align}
\eta_{+}({\bf X})
=
\frac{|a|}{\sqrt{a^2-b^2}}
\left(
\mathbb I+\vec{\beta}\cdot\vec{\sigma}
\right),
\label{eq:eta_2x2}
\end{align}
where the vector
\begin{align}
\vec{\beta}
=
\frac{b}{a}
\left(
\sin\delta\,\hat{e}_{\theta}
-
\cos\delta\,\hat{e}_{\varphi}
\right),
\end{align}
encodes the components of the metric operator along the tangent directions of the Bloch sphere. One can verify that $\eta_+$ is strictly positive definite, thereby providing the necessary mathematical foundation for constructing the quasi-Hermitian inner product, similarity transformations, and the geometric structure of mixed states.

Under the above quasi-Hermitian conditions, the right eigenstates of $H_{2\times2}({\bf X})$ can be chosen as
\begin{align}
|\psi_{\pm}\rangle
=
\frac{1}{\mathcal N_{\pm}}
\begin{pmatrix}
\mathrm{e}^{-\mi\phi}
\left(
a\sin\theta
+\mi b\cos\delta\cos\theta
+b\sin\delta
\right)
\\[4pt]
- a\cos\theta
+\mi b\cos\delta\sin\theta
\pm\sqrt{a^2-b^2}
\end{pmatrix},
\label{eq:PT_eigenstates}
\end{align}
with normalization factors
\begin{align}
\mathcal N_{\pm}^2
=
2|a|\sqrt{a^2-b^2}
\left[
1+\frac{b}{a}\sin\delta\sin\theta
\mp
\frac{\sqrt{a^2-b^2}}{a}\cos\theta
\right].
\end{align}
Compared to a generic Hermitian $2\times2$ Hamiltonian, this $\mathcal{PT}$-symmetric model introduces two additional independent parameters, offering a minimal yet versatile platform for exploring the richness of geometric structures in non-Hermitian systems.

\subsubsection{Uhlmann Connection and Phase on an Equatorial Closed Loop}

To simplify analytic calculations, we substitute the spectral decomposition of the density matrix, Eq.~\eqref{DMEB}, into the general expression for the Uhlmann connection, Eq.~\eqref{eq:qh_A_explicit_final}. This yields an alternative form
\begin{align}
\mathcal{A}_{\text{U}}^{\eta_+}
=
-\sum_{i\neq j}
\frac{(\sqrt{p_i}-\sqrt{p_j})^2}{p_i+p_j}
\,|\Psi_i\rangle\langle\Phi_i|\,
\dif|\Psi_j\rangle\langle\Phi_j|,
\label{eq:AU_twolevel_general}
\end{align}
where $p_i$ are the eigenvalues of $\rho$ and $\{|\Psi_i\rangle,\langle\Phi_i|\}$ denotes the corresponding biorthogonal eigenbasis. This expression contains only off-diagonal terms, significantly simplifying analytic calculations for two-level systems.

We consider a closed loop on the equator of the Bloch sphere in parameter space, i.e., we fix $\theta=\pi/2$ and let $\phi:0\to2\pi$, while keeping all other parameters constant. The system is taken to be in the quasi-Hermitian equilibrium state $\rho=\me^{-\beta H}/Z$, with $\varepsilon=0$, $\delta=0$, and $a,b$ as general real parameters. In the quasi-Hermitian regime $a^2>b^2$, the energy gap is $\Delta=2\sqrt{a^2-b^2}$, and the density matrix eigenvalues are $p_\pm=\frac12\left(1\mp\tanh(\frac{\beta\Delta}{2})\right)$.

Substituting these results into Eq.~\eqref{eq:AU_twolevel_general} and introducing
$\kappa=(\sqrt{p_+}-\sqrt{p_-})^2 =1-\text{sech}(\frac{\beta\Delta}{2})$,
the quasi-Hermitian Uhlmann connection along the equatorial path becomes
\begin{align}
\mathcal{A}_{\text{U}}^{\eta_+}
=
\kappa
\left[
\frac{2a^2}{\Delta^2}\,\mi\,\sigma_z
-
\frac{2ab}{\Delta^2}
\me^{-\mi\phi\sigma_z /2}\sigma_x \me^{\mi\phi\sigma_z /2}
\right]
\,\dif\phi.
\label{eq:AU_eta_equator_general_ab}
\end{align}
The Uhlmann holonomy after one full loop around the equator can be computed explicitly \cite{Galindo21, Bohm2003GeometricPhase}:
\begin{align}\label{Uhlholonomy}
U(2\pi\Omega)= (-1)^{\Omega}\me^{2\pi\Omega\Big[\mi\Big(\frac{1}{2}-\frac{2a^2\kappa}{\Delta^2}\Big)\sigma_{z}+\frac{2ab\kappa}{\Delta^2}\sigma_x\Big]},
\end{align}
where $\Omega$ denotes the winding number around the equatorial loop. A detailed derivation is provided in Appendix~\ref{app:equator_holonomy}.

Substituting the above holonomy into Eq.~\eqref{QHUPG}, we obtain the Uhlmann amplitude
\begin{align}
\mathcal{G}(\beta)
=
(-1)^{\Omega}
\left[
\cos\Theta
-
\frac{b}{\Delta}
\tanh\!\left(\frac{\Delta\beta}{2}\right)
\frac{\sin\Theta}{\gamma}
\right],
\label{GUequator}
\end{align}
where
\begin{align}
\Theta = 2\pi\Omega\,\gamma,
\quad
\gamma=
\sqrt{
\left(
\frac{1}{2}
-
\frac{2a^2}{\Delta^2}\kappa
\right)^2
-
\left(
\frac{2ab}{\Delta^2}\kappa
\right)^2
}.
\end{align}
The effect of the non-Hermitian parameter $b$ can be read directly from this expression. Besides changing the spectral gap $\Delta$ and hence the thermal weights, $b$ also enters the metric $\eta_+(\lambda)$ that determines the physical overlap between neighboring purifications. Varying $b$ therefore affects both the statistical weights of the mixed state and the local Hilbert-space geometry along the loop. Consequently, the zeros of $\mathcal{G}(\beta)$ reflect the combined influence of thermal mixing and the parameter-dependent metric.

The Uhlmann phase is then obtained from $\theta_{\text{U}}^{\eta_+}=\arg\mathcal{G}(\beta)$. When the non-Hermitian parameter $b=0$, this result reduces to the Hermitian case \cite{PhysRevA.104.023303}. Although $\gamma$ may become purely imaginary in certain parameter regimes, the Uhlmann amplitude $\mathcal{G}(\beta)$ remains real. Indeed, if $\gamma=\mi\mu$ with $\mu\in\mathbb{R}$, then $\Theta=2\pi\Omega \mi\mu$, and
\begin{align}
\cos\Theta
=
\cosh(2\pi\Omega\mu),
\quad
\frac{\sin\Theta}{\gamma}
=
\frac{\sinh(2\pi\Omega\mu)}{\mu},
\end{align}
which are both real quantities. Hence $\mathcal{G}(\beta)\in\mathbb{R}$ in both the oscillatory regime $\gamma^2>0$ and the hyperbolic regime $\gamma^2<0$. Therefore, the Uhlmann phase along the equatorial loop is completely determined by the sign of $\mathcal{G}(\beta)$. When $\mathcal{G}(\beta)$ crosses zero as temperature varies, the Uhlmann phase undergoes a $\pi$-jump, signaling a geometric phase transition.

In the infinite-temperature limit $\beta\to0$, we have $\tanh(\frac{\Delta\beta}{2})\to0$ and $\kappa\to0$, which leads to $\gamma\to\frac{1}{2}$ and $\Theta\to\pi\Omega$. Since $\Omega$ is an integer, the Uhlmann phase satisfies $\theta_{\text{U}}^{\eta_+}\to0\;(\mathrm{mod}\,2\pi)$. This indicates that thermal fluctuations wash out the geometric structure at high temperatures, leaving the system in a trivial phase.

As temperature decreases, $\mathcal{G}(\beta)$ gradually deviates from its high-temperature limit and may cross zero at critical temperatures, inducing $\pi$-jumps in the Uhlmann phase. These transitions reflect the nonanalytic response of the mixed-state geometric phase to temperature and signal finite-temperature geometric phase transitions. Thus, temperature acts as a control parameter that reshapes the geometric phase structure, allowing the system to enter geometrically nontrivial sectors over certain finite-temperature intervals.

In the zero-temperature limit $\beta\to\infty$, we obtain $\kappa\to1$, $\gamma(\beta)\to\gamma_0$, and $\tanh(\Delta\beta/2)\to1$. The Uhlmann amplitude therefore takes the limiting form
\begin{align}
\mathcal G(\infty)
=
(-1)^{\Omega}
\left[
\cos(2\pi\Omega\gamma_0)
-
\frac{b}{\Delta\gamma_0}
\sin(2\pi\Omega\gamma_0)
\right],
\end{align}
where
\begin{align}
\gamma_0^2
=
\left(
\frac{1}{2}
-
\frac{2a^2}{\Delta^2}
\right)^2
-
\left(
\frac{2ab}{\Delta^2}
\right)^2
=
-\frac{b^2}{\Delta^2}.
\end{align}
For $b\neq0$, $\gamma_0$ is purely imaginary. Since $\mathcal G(\infty)$ is unchanged under $\gamma_0\to-\gamma_0$, we set $\gamma_0=\mi b/\Delta$ for convenience. The trigonometric functions in $\mathcal G(\infty)$ then become hyperbolic functions, yielding
\begin{align}
\cos(2\pi\Omega\gamma_0)
&=
\cosh\left(\frac{2\pi\Omega b}{\Delta}\right), \notag \\
\sin(2\pi\Omega\gamma_0)
&=
\mi\sinh\left(\frac{2\pi\Omega b}{\Delta}\right).
\end{align}
Substituting these relations into the zero-temperature Uhlmann amplitude gives
\begin{align}
\mathcal G(\infty)
=
(-1)^{\Omega}
\exp\left(
-\frac{2\pi\Omega b}{\Delta}
\right).
\end{align}
Thus, at zero temperature the non-Hermitian parameter $b$ merely rescales the magnitude of the Uhlmann amplitude by a real exponential factor. Since $\exp(-\frac{2\pi\Omega b}{\Delta})>0$, the sign of $\mathcal G(\infty)$, and hence the Uhlmann phase, is determined solely by the parity of the winding number $\Omega$. This agrees with the Hermitian spin-$1/2$ result, where the zero-temperature geometric phase is governed by the sign of $\cos(\pi\Omega)$ \cite{Galindo21,PhysRevA.104.023303}.

We now turn to the finite-temperature case. Here, the Uhlmann phase is no longer fixed solely by the winding number $\Omega$, but is jointly controlled by the temperature $T$ and the quasi-Hermitian coefficient $b$. This dependence enters through the thermal weight factor $\kappa(\beta)$, the effective parameter $\gamma(\beta)$, and the explicit $b/\Delta$ term in the Uhlmann amplitude, shifting and deforming the zeros of $\mathcal G(\beta)$ in the $(b,T)$ plane and thereby producing jumps of the Uhlmann phase.

\begin{figure}[th]
\centering
\includegraphics[width=3.3in]{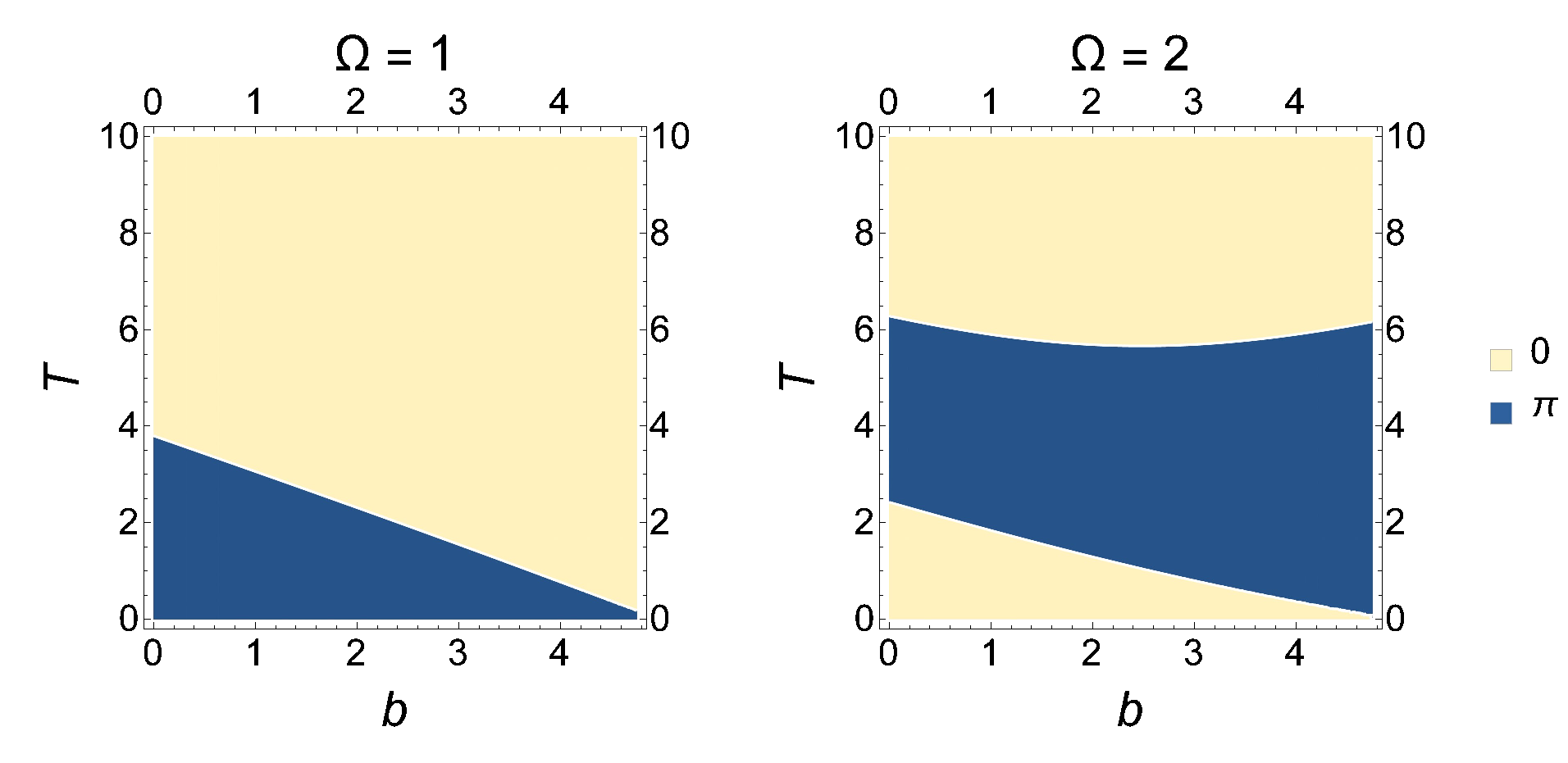}
\caption{Uhlmann phase $\theta_{\text{U}}^{\eta_+}$ for the quasi-Hermitian two-level model along an equatorial closed loop as a function of temperature $T$ and the quasi-Hermitian coefficient $b$. Left and right panels correspond to winding numbers $\Omega=1$ and $\Omega=2$, respectively. Yellow: $\theta_{\text{U}}^{\eta_+}=0$. Dark blue: $\theta_{\text{U}}^{\eta_+}=\pi$. These two regions correspond to distinct Uhlmann-phase sectors. Parameters: $\varepsilon=0$, $a=5$, $\delta=0$.}
\label{fig2}
\end{figure}

Fig.~\ref{fig2} displays the Uhlmann phase $\theta_{\text{U}}^{\eta_+}$ for the quasi-Hermitian two-level model along an equatorial closed loop as a function of temperature $T$ and the quasi-Hermitian coefficient $b$. The left and right panels correspond to winding numbers $\Omega=1$ and $\Omega=2$, respectively. Yellow regions indicate $\theta_{\text{U}}^{\eta_+}=0$, while dark blue regions indicate $\theta_{\text{U}}^{\eta_+}=\pi$. At infinite temperature, $\kappa(\beta)\to0$ and $\gamma(\beta)\to1/2$, so the Uhlmann phase always tends to zero, showing that thermal fluctuations wash out the geometric phase. In the zero-temperature limit, by contrast, the sign of $\mathcal G(\infty)$ is fixed by $(-1)^\Omega$, in agreement with the parity criterion discussed above.

\begin{figure}[th]
\centering
\includegraphics[width=3.3in]{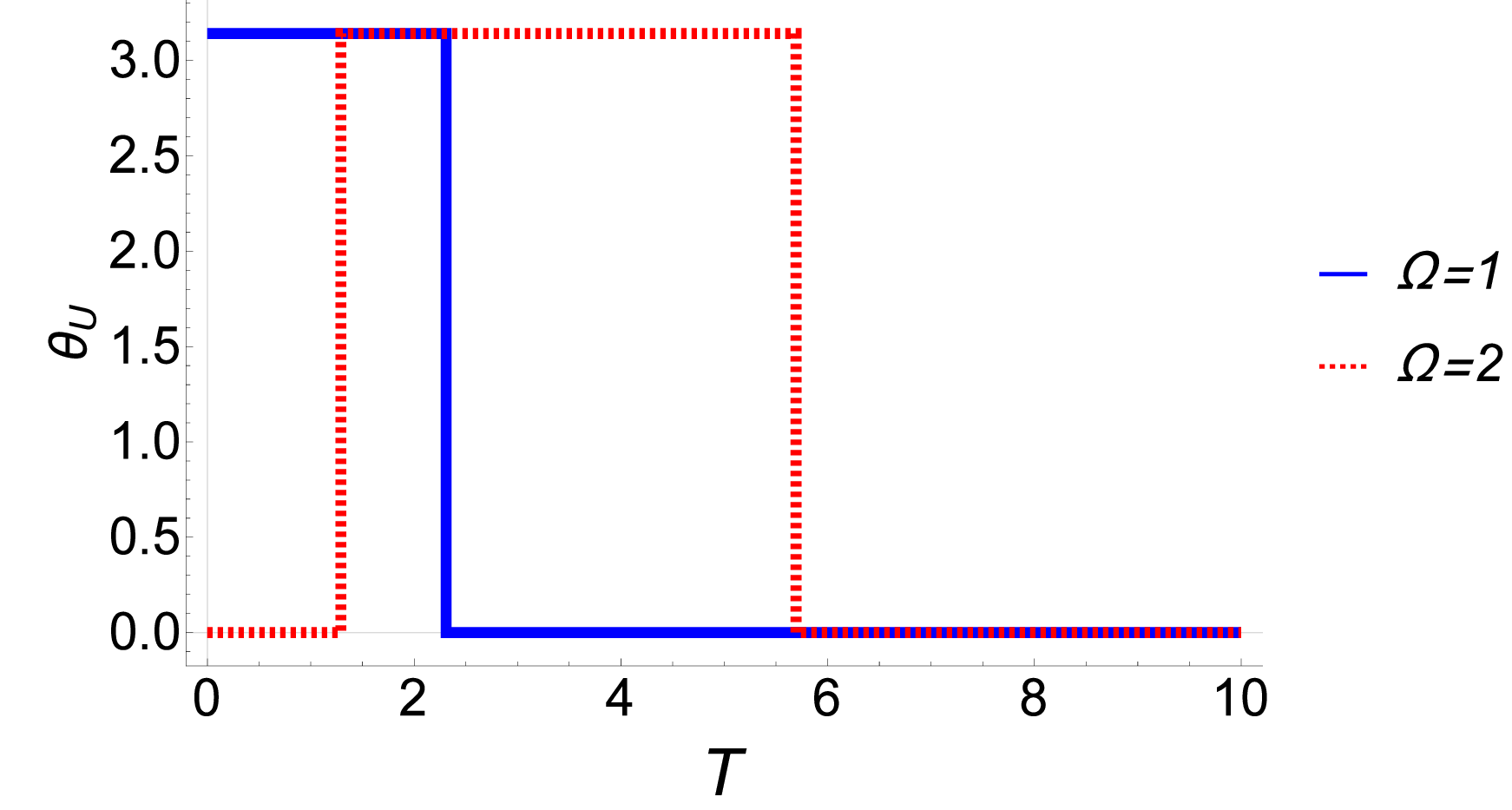}
\caption{Uhlmann phase $\theta^{\eta_{+}}_{U}$ for the quasi-Hermitian two-level model along an equatorial closed loop as a function of temperature $T$. Blue solid and red dashed lines correspond to winding numbers $\Omega=1$ and $2$, respectively. Parameters: $\varepsilon=0$, $a=5$, $b=2$, $\delta=0$.}
\label{Fig3}
\end{figure}

At finite temperature, the zeros of $\mathcal G(\beta)$ give rise to distinct transition patterns for different winding numbers. For $\Omega=1$, the nontrivial $\pi$ sector appears in the low-temperature region and is separated from the trivial sector by a single zero line of the Uhlmann amplitude. As $b$ increases, this transition line shifts toward lower temperatures, indicating that the finite-temperature survival window of the $\pi$ phase is gradually reduced. For $\Omega=2$, the phase is trivial both at zero and infinite temperature, while a nontrivial $\pi$ sector emerges at intermediate temperatures. This sector is bounded by two zero lines of $\mathcal G(\beta)$, and its shape is significantly deformed by the quasi-Hermitian coefficient $b$. To further reveal the influence of temperature on the geometric phase structure in quasi-Hermitian systems, Fig.~\ref{Fig3} shows the Uhlmann phase as a function of temperature for fixed parameters $a=5$ and $b=2$, providing a clearer view of the transition behavior.

Therefore, quasi-Hermiticity significantly alters the finite-temperature response of the Uhlmann phase. Compared with the Hermitian spin-$\frac{1}{2}$ case, where the transition pattern is fixed by the winding number $\Omega$ \cite{Galindo21,PhysRevA.104.023303}, the quasi-Hermitian system exhibits a tunable zero-crossing structure of the Uhlmann amplitude. The parameter-dependent metric shifts and deforms the zero lines of $\mathcal G(\beta)$, thereby redistributing the temperature intervals in which nontrivial Uhlmann phases occur. This shows that the modified Hilbert-space structure does not simply change the winding-number criterion uniformly; rather, it changes how the geometric phase responds to thermal fluctuations and to variations of the quasi-Hermitian coefficient. Hence, quasi-Hermiticity provides a mechanism for tuning geometric phase transitions in mixed states at finite temperature.

\section{Interferometric access to the quasi-Hermitian Uhlmann phase}
\label{sec:experimental_realization}

The quasi-Hermitian Uhlmann phase developed in this work is, in principle, experimentally accessible. By embedding the purification formalism into an enlarged Hilbert space, the associated geometric information can be extracted through measurements of the overlap between purified states. Crucially, the quasi-Hermitian Hilbert-Schmidt inner product that characterizes the geometric amplitude can be reformulated, within an enlarged purified-state framework, as a Loschmidt amplitude between two pure states, thereby providing a clear operational route for experimental detection.

Starting from the quasi-Hermitian density matrix $\rho(\lambda)$ introduced in Eq.~\eqref{DMEB}, whose spectral decomposition is given in the biorthogonal basis $\{|\psi_n\rangle,\langle\phi_n|\}$, the corresponding amplitude operator $W=\sqrt{\rho}\,\mathcal U$ can be written in spectral form as
\begin{align}
W = \sum_{n} \sqrt{p_n}\, |\psi_n\rangle \langle\phi_n|\, \mathcal U .
\end{align}

To obtain an experimentally accessible formulation, we recast the amplitude operator in its vectorized form, mapping it onto a purified state in an enlarged Hilbert space $\mathcal H_S \otimes \mathcal H_A$~\cite{ourPRB20, OurPRB20b},
\begin{align}
|W\rangle = \sum_{n} \sqrt{p_n}\, |\psi_n\rangle_S \otimes \mathcal U^{T}|\phi_n\rangle_A .
\end{align}

This construction establishes an isomorphism between the amplitude space and the tensor-product Hilbert space of the system and an ancilla. Under this mapping, the quasi-Hermitian Hilbert-Schmidt product becomes
\begin{align}
\mathrm{Tr}_{\eta_+}\!\left(W_1^{\ddagger} W_2\right) = \langle W_1 | W_2 \rangle_{\eta_+ \otimes \eta_+^{-1}},
\end{align}
with a detailed proof provided in Appendix~\ref{app:purification_proof}.

In practice, the metric dependence is introduced through the preparation and comparison of purified states. The right eigenvectors are encoded in the system, while the corresponding left eigenvectors are encoded in the ancilla. The metric-weighted overlap is then represented through the biorthogonal relation $\langle \phi_m(\lambda)|\psi_n(\lambda)\rangle=\delta_{mn}$, rather than by measuring the metric operator as an independent observable. Consequently, the geometric amplitude admits an equivalent pure-state representation
\begin{align}
\mathcal G = \langle W(0) | W(\tau) \rangle .
\end{align}

Thus, the Hilbert-Schmidt inner product between initial and final purifications can be interpreted as the Loschmidt amplitude, i.e., the Uhlmann amplitude between the corresponding purified states in the enlarged Hilbert space~\cite{Heyl_2018, OurPRB20b, PhysRevB.107.184311}. The quasi-Hermitian Uhlmann phase $\theta_{\text{U}}^{\eta_+} = \arg \mathcal G$ is therefore directly encoded in the complex overlap of these two pure states.

To access the complex amplitude $\mathcal G$, one can implement an ancilla-assisted interferometric protocol~\cite{Viyuela2018TopologicalUhlmann, Mastandrea_2026}. By introducing an additional control qubit and performing controlled state preparation for the initial and final purifications, the joint system is prepared in the entangled superposition
\begin{align}
\frac{1}{\sqrt{2}} \big( |0\rangle_c \otimes |W(0)\rangle + |1\rangle_c \otimes |W(\tau)\rangle \big).
\end{align}
Measurements of $\sigma_x$ and $\sigma_y$ on the control qubit then yield the real and imaginary parts of $\mathcal G$, respectively, enabling direct measurement of the Uhlmann phase.

The quasi-unitary evolution enters the protocol through the Uhlmann parallel transport of the purification amplitude. The initial and final purifications are related by \begin{align}
|W(\tau)\rangle &= U(\tau)|W(0)\rangle,  \notag \\
 U(\tau) &= \mathcal P \exp\!\left[ -\int_{\lambda(0)}^{\lambda(\tau)} \mathcal A_{\text{U}}^{\eta_+} \right].
\end{align}
For Hermitian systems, Ref.~\cite{Mastandrea_2026} has demonstrated the simulation and measurement of the Uhlmann phase on IBM quantum computers by discretizing the parameter path and implementing the corresponding parallel-transport steps with standard quantum gates. A similar strategy can be applied to the quasi-Hermitian case: by discretizing the path and applying gate sequences that realize the quasi-Hermitian Uhlmann connection $\mathcal A_{\text{U}}^{\eta_+}$, the ordered product of these gates reproduces the path-ordered holonomy $U(\tau)$ acting on the prepared purification. Thus, the quasi-Hermitian Uhlmann transport can be probed experimentally via gate-based simulations along the chosen path, without requiring the construction of new elementary quasi-unitary gates in the ordinary Dirac Hilbert space.

For the two-level models considered, experimental implementation requires minimal resources: a system qubit, an ancilla for purification, and a control qubit for interferometric readout.
The relevant operations  can be decomposed into elementary gates. Specifically, preparing $|W(0)\rangle$ and $|W(\tau)\rangle$ uses only parameterized single-qubit rotations, while the controlled operations can be realized with standard CNOT gates. These requirements are well within the capabilities of current quantum processors, such as IBM devices, where high-fidelity single- and two-qubit gates are routinely available. For instance, the controlled preparation of $|W(\tau)\rangle$ can be implemented by first preparing $|W(0)\rangle$ and then applying a parameterized circuit that implements $U(\tau)$ conditional on the control qubit.

Since $\mathcal G$ is directly accessible within this interferometric framework, finite-temperature geometric phase transitions acquire a clear experimental signature. The transition is characterized by the vanishing of the Loschmidt amplitude, $\mathcal G = 0$, which manifests experimentally as the loss of interference contrast in the control qubit. By systematically varying the effective temperature and tracking the measured overlap signal, the critical points can be directly identified.
In the purified-state protocol, this variation is realized at the level of state preparation: changing $\beta$ amounts to changing the thermal amplitudes $\sqrt{p_n(\beta)}$, without physically heating the device.

This provides a concrete and experimentally viable pathway for probing quasi-Hermitian mixed-state geometric phases and their temperature-driven transitions on quantum information platforms.

\section{Conclusion}
In this work, we have established a comprehensive theoretical framework for the Uhlmann phase in quasi-Hermitian quantum systems. Recognizing that the standard purification $\rho = WW^\dagger$ is incompatible with non-Hermitian structures, we introduced a generalized amplitude decomposition $\rho = WW^{\ddagger}$ based on the physical inner product defined by the positive-definite metric operator $\eta_+$. This construction naturally yields a quasi-unitary gauge group $U_{\eta_+}(N)$ and a parallel transport condition that minimizes the $\eta_+$-weighted Hilbert-Schmidt distance. The resulting Uhlmann connection and its holonomy provide a gauge-invariant characterization of the geometric phase for mixed states in systems with parameter-dependent inner products.

Our analysis reveals that the parameter-dependent metric introduces an additional geometric contribution to the parallel transport of quasi-Hermitian mixed states. This contribution modifies the Uhlmann connection and affects the finite-temperature behavior of the Uhlmann phase, leading to a response distinct from that of Hermitian mixed states. Using exactly solvable two-level models, we demonstrated that temperature can drive transitions between distinct Uhlmann-phase sectors. The winding number $\Omega$ determines the basic transition pattern, while the quasi-Hermitian coefficient shifts and deforms the zero lines of the Uhlmann amplitude in the finite-temperature phase diagram. In the $\mathcal{PT}$-symmetric model, the zero-temperature parity criterion remains unchanged, but the finite-temperature regions supporting nontrivial Uhlmann phases are reshaped by the quasi-Hermitian coefficient $b$.

Furthermore, we have shown that the quasi-Hermitian Uhlmann phase is experimentally accessible via an ancilla-assisted interferometric protocol, where the geometric amplitude is encoded as a measurable Loschmidt overlap between purified states. The required quantum resources are minimal and well within the capabilities of existing quantum information platforms, providing a practical pathway for probing finite-temperature geometric phase transitions in quasi-Hermitian systems. In particular, the framework connects naturally to the mixed-state quantum geometric tensor, providing a unified description of both local and global geometry in quasi-Hermitian systems.

These findings highlight the intricate interplay between thermal fluctuations, the quasi-Hermitian metric structure, and geometric phases. The theory developed here provides a foundation for investigating finite-temperature geometric structure in a broad class of non-Hermitian systems admitting a quasi-Hermitian description, including open quantum systems, optical platforms with gain and loss, and correlated materials described by effective non-Hermitian Hamiltonians. Future directions include extending this framework to higher-dimensional parameter spaces and interacting many-body systems, elucidating its connections with quantum geometric tensors and non-Hermitian topological invariants, and systematically exploring the geometric role of parameter-dependent metrics in more general pseudo-Hermitian and
genuinely complex-spectral settings. Advancing along these lines promises deeper insight into the structure of non-Hermitian quantum geometry and its interplay with thermodynamic phenomena.
\section{Acknowledgements}
H. G. was supported by the Quantum Science and Technology-National
Science and Technology Major Project (Grant No. 2021ZD0301904) and the National Natural Science
Foundation of China (Grant No. 12447216). X.-Y. H. was supported by the National Natural Science
Foundation of China (Grant No. 12405008).
\appendix
\section{Derivation of the Uhlmann Parallel Transport Condition}
\label{app:derivation_parallel_condition}

In quasi-Hermitian systems, the density matrix can be written as $\rho = W W^{\ddagger}$, where the purification (amplitude) $W$ possesses a gauge redundancy $W \sim W U$ with $U \in U_{\eta_+}(N)$ satisfying $U^{\ddagger}U = UU^{\ddagger} = \mathbb{I}$. To select the most natural lift among the gauge degrees of freedom, we introduce the physical Hilbert--Schmidt distance on the purification space:
\begin{align}
D_{\mathrm{HS}}^2(W_1,W_2) := \mathrm{Tr}_{\eta_+}\!\left[(W_1-W_2)^{\ddagger}(W_1-W_2)\right],
\label{eq:HS_def_app2}
\end{align}
where $\mathrm{Tr}_{\eta_+}(X) :=  \sum_n \langle \Phi_n | X | \Psi_n \rangle$. Given two neighboring parameter points $\lambda$ and $\lambda+\dif\lambda$, we denote $W_1 := W(\lambda)$ and $W_2 := W(\lambda+\dif\lambda)$. Since the purification at the second point can undergo a gauge transformation, we minimize the function over all $U \in U_{\eta_+}(N)$:
\begin{align}
F(U) &:= D_{\mathrm{HS}}^2\!\bigl(W_1,\,W_2U\bigr) \notag \\
&= \mathrm{Tr}_{\eta_+}\!\left[(W_1-W_2U)^{\ddagger}(W_1-W_2U)\right].
\label{eq:F_V_def}
\end{align}

Expanding \eqref{eq:F_V_def} and using $U^{\ddagger}U = \mathbb{I}$ yields
\begin{align}
F(U) &= \mathrm{Tr}_{\eta_+}(W_1^{\ddagger}W_1) + \mathrm{Tr}_{\eta_+}(U^{\ddagger}W_2^{\ddagger}W_2U) \notag \\
&\quad -\mathrm{Tr}_{\eta_+}(W_1^{\ddagger}W_2U) -\mathrm{Tr}_{\eta_+}(U^{\ddagger}W_2^{\ddagger}W_1) \notag\\
&= \mathrm{Tr}_{\eta_+}(W_1^{\ddagger}W_1) + \mathrm{Tr}_{\eta_+}(W_2^{\ddagger}W_2) \notag \\
&\quad -2\,\mathrm{Re}\,\mathrm{Tr}_{\eta_+}(W_1^{\ddagger}W_2U),
\label{eq:F_expand}
\end{align}
where we used $\mathrm{Tr}_{\eta_+}(U^{\ddagger}X U) = \mathrm{Tr}_{\eta_+}(X)$ (since $U$ is $\eta_+$-unitary). Minimizing $F(U)$ is therefore equivalent to maximizing
\begin{align}
G(U) := \mathrm{Re}\,\mathrm{Tr}_{\eta_+}(W_1^{\ddagger}W_2U).
\label{eq:G_def}
\end{align}

To find the extremum condition, consider an infinitesimal variation of $U$. Since $U \in U_{\eta_+}(N)$, we write
\begin{align}
U \ \rightarrow\ U\,\me^{\epsilon K},\qquad K^{\ddagger} = -K,\qquad \epsilon\in\mathbb{R},\ |\epsilon|\ll 1,
\end{align}
so that $\delta U = \epsilon\,U K$ and $\delta U^{\ddagger} = -\epsilon\,K U^{\ddagger}$. The first variation of \eqref{eq:G_def} gives
\begin{align}
\delta G = \epsilon\,\mathrm{Re}\,\mathrm{Tr}_{\eta_+}\!\left(W_1^{\ddagger}W_2\,U K\right).
\label{eq:deltaG}
\end{align}
The extremum requires $\delta G = 0$ for all anti-self-adjoint $K^{\ddagger} = -K$, i.e.,
\begin{align}
\mathrm{Re}\,\mathrm{Tr}_{\eta_+}\!\left(AK\right) = 0,\qquad A := W_1^{\ddagger}W_2U.
\label{eq:AK_cond}
\end{align}
Any operator can be decomposed into $\ddagger$-self-adjoint and $\ddagger$-anti-self-adjoint parts: $A = A_{\mathrm{h}} + A_{\mathrm{a}}$ with $A_{\mathrm{h}} = \frac12(A+A^{\ddagger})$ and $A_{\mathrm{a}} = \frac12(A-A^{\ddagger})$. For any $K^{\ddagger} = -K$, $\mathrm{Tr}_{\eta_+}(A_{\mathrm{h}}K)$ is purely imaginary while $\mathrm{Tr}_{\eta_+}(A_{\mathrm{a}}K)$ is purely real. Hence condition \eqref{eq:AK_cond} holds for all anti-self-adjoint $K$ if and only if
\begin{align}
A_{\mathrm{a}} = \frac12(A-A^{\ddagger}) = 0 \quad \Longleftrightarrow\quad A = A^{\ddagger}.
\label{eq:A_selfadj}
\end{align}
Thus, at the extremal gauge we must have
\begin{align}
\bigl(W_1^{\ddagger}W_2U\bigr)^{\ddagger} = W_1^{\ddagger}W_2U,
\label{eq:stationary_selfadj}
\end{align}
i.e., $W_1^{\ddagger}W_2U$ is $\ddagger$-self-adjoint. Moreover, in the full-rank case, the solution that maximizes \eqref{eq:G_def} corresponds to the $\ddagger$-polar decomposition of $W_1^{\ddagger}W_2U$, with the optimal choice making this operator positive definite, often written as
\begin{align}
W_1^{\ddagger}W_2U > 0.
\label{eq:positive_condition_app2}
\end{align}
This condition is the discrete version of the Uhlmann parallel condition commonly found in the literature.

When the metric operator is parameter independent, the above condition can be equivalently written, after choosing the parallel gauge and relabeling $W_2U$ as $W_2$, as
\begin{align}
    W_1^\ddagger W_2=W_2^\ddagger W_1>0 .
    \label{eq:fixed_metric_parallel_discrete_app}
\end{align}
This condition states that the forward and backward overlaps are related by the same fixed quasi-Hermitian adjoint.

For a parameter-dependent metric, however, the quasi-Hermitian adjoint depends on the endpoint, \begin{align} X^{\ddagger_\lambda} = \eta_+^{-1}(\lambda)X^\dagger\eta_+(\lambda). \end{align} Thus the two neighboring points $\lambda$ and $\lambda+\dif\lambda$ generally carry different adjoint operations. The natural path-compatible extension of Eq.~\eqref{eq:fixed_metric_parallel_discrete_app} is therefore
\begin{align}
W^{\ddagger_\lambda}(\lambda)W(\lambda+\dif\lambda) = W^{\ddagger_{\lambda+\dif\lambda}}(\lambda+\dif\lambda)W(\lambda) >0 . \label{eq:path_compatible_discrete_parallel_app}
\end{align}

Since the metric varies along the path, the neighboring amplitudes are compared with respect to different endpoint physical inner products. The condition \eqref{eq:path_compatible_discrete_parallel_app} therefore compares the corresponding physical overlaps at the two endpoints, rather than imposing a minimization in an artificially fixed Hilbert--Schmidt space.

Taking the continuum limit, we write $W(\lambda+\dif\lambda)=W(\lambda)+\dif W(\lambda)$ and choose the extremal gauge such that $U=\mathbb I+\mathcal O(\dif\lambda)$. Since the adjoint itself is parameter dependent, the neighboring adjoint amplitude has the first-order expansion
\begin{align} W^{\ddagger_{\lambda+\dif\lambda}}(\lambda+\dif\lambda) = W^{\ddagger_\lambda}(\lambda) + \dif\!\left[ W^{\ddagger_\lambda}(\lambda) \right] + \mathcal O(\dif\lambda^2).
\end{align}
Substituting these expansions into Eq.~\eqref{eq:path_compatible_discrete_parallel_app} and keeping only first-order terms gives
\begin{align}
    W^{\ddagger_\lambda}(\lambda)\dif W(\lambda)
    =
    \dif\!\left[
    W^{\ddagger_\lambda}(\lambda)
    \right]W(\lambda).
    \label{eq:path_uhlmann_parallel_final_app}
\end{align}
Equivalently, suppressing the explicit parameter labels, the path-compatible Uhlmann parallel transport condition is
\begin{align}
    W^\ddagger \dif W
    =
    \dif(W^\ddagger)W .
    \label{eq:path_uhlmann_parallel_simple_app}
\end{align}
Here $\dif(W^\ddagger)$ denotes the full path differential of the quasi-Hermitian adjoint, including the variation of both $W$ and $\eta_+$. Since $W^\ddagger=\eta_+^{-1}W^\dagger\eta_+$, this full differential satisfies
\begin{align}
\dif(W^\ddagger)
=
(\dif W)^\ddagger
-
\Gamma W^\ddagger
+
W^\ddagger\Gamma,
\quad
\Gamma=\eta_+^{-1}\dif\eta_+ .
\end{align}
Hence $\dif(W^\ddagger)$ and $(\dif W)^\ddagger$ differ by the metric-induced connection terms, and they coincide only when the metric is fixed along the path.

\section{Derivation of the Quasi-Hermitian Uhlmann Connection}
\label{app:derivation_uhlmann_connection}

Here we provide a detailed derivation of the quasi-Hermitian Uhlmann connection directly from the quasi-Hermitian parallel evolution condition, demonstrating how the connection is uniquely determined by the parallel transport condition.
Consider a parameter path $\lambda \mapsto \rho(\lambda)$ with the corresponding purification representation
\begin{align}
\rho(\lambda) = W(\lambda) W^{\ddagger}(\lambda),
\end{align}
where $\ddagger$ denotes the adjoint operation defined with respect to the quasi-Hermitian physical inner product. The Uhlmann parallel evolution condition for mixed states in quasi-Hermitian systems reads
\begin{align}
W^{\ddagger}\dot W
=
\dot{(W^{\ddagger})}W,
\end{align}
with $\dot{(\,)} \equiv \dif/\dif\lambda$.
In the full-rank case, we employ the gauge decomposition introduced in the main text:
\begin{align}
W(\lambda) = \sqrt{\rho(\lambda)}\,U(\lambda),\qquad U^{\ddagger}U = \mathbb{I},
\end{align}
where $U(\lambda)$ takes values in the quasi-unitary group $U_{\eta_+}(N)$. Differentiating $W$ gives
\begin{align}
\dot W = \dot{\sqrt{\rho}}\,U + \sqrt{\rho}\,\dot U .
\end{align}
Using the property $(AB)^{\ddagger} = B^{\ddagger}A^{\ddagger}$ and the constraint $U^{\ddagger}U = \mathbb{I}$, we compute
\begin{align}
W^{\ddagger}\dot W
&= U^{\ddagger}\sqrt{\rho}\left( \dot{\sqrt{\rho}}\,U + \sqrt{\rho}\,\dot U \right) \notag\\
&= U^{\ddagger}\!\left(\sqrt{\rho}\,\dot{\sqrt{\rho}}\right)\!U + U^{\ddagger}\rho\,\dot U,
\end{align}
and
\begin{align}
\dot{(W^{\ddagger})}W
&=
\left[
\dot{(U^{\ddagger})}\sqrt{\rho}
+
U^{\ddagger}\dot{\sqrt{\rho}}
\right]\sqrt{\rho}\,U
\notag\\
&=
\dot{(U^{\ddagger})}\rho\,U
+
U^{\ddagger}\!\left(\dot{\sqrt{\rho}}\,\sqrt{\rho}\right)\!U .
\end{align}

Substituting these results into the parallel evolution condition and multiplying on the left by $U$ and on the right by $U^{\ddagger}$, we obtain
\begin{align}
\sqrt{\rho}\,\dot{\sqrt{\rho}}
+
\rho\,\dot U U^{\ddagger}
=
\dot{\sqrt{\rho}}\,\sqrt{\rho}
+
U\dot{(U^{\ddagger})}\rho .
\label{eq:app_intermediate}
\end{align}
Differentiating the identity $U^{\ddagger}U = \mathbb{I}$ yields
\begin{align}
\dot{(U^{\ddagger})}U
+
U^{\ddagger}\dot U
=
0.
\end{align}
Equivalently,
\begin{align}
U\dot{(U^{\ddagger})}
=
-\dot U U^{\ddagger}.
\end{align}

Introducing the definition of the quasi-Hermitian Uhlmann connection $\mathcal{A}_{\text{U}}^{\eta_+} := -\dif U U^{\ddagger}$,
and substituting into \eqref{eq:app_intermediate}, we obtain
\begin{align}
\sqrt{\rho}\,\dif\sqrt{\rho} - \rho\,\mathcal{A}_{\text{U}}^{\eta_+} = \dif\sqrt{\rho}\,\sqrt{\rho} + \mathcal{A}_{\text{U}}^{\eta_+}\rho .
\end{align}
Rearranging gives the Sylvester-type equation satisfied by $\mathcal{A}_{\text{U}}^{\eta_+}$:
\begin{align}
\rho\,\mathcal{A}_{\text{U}}^{\eta_+} + \mathcal{A}_{\text{U}}^{\eta_+}\rho = -\bigl[\dif\sqrt{\rho},\sqrt{\rho}\bigr].
\end{align}
Thus, the quasi-Hermitian Uhlmann connection is uniquely determined by the parallel evolution condition, its form depending solely on the variation of the mixed-state density matrix along the parameter path. The geometric phase for quasi-Hermitian mixed states discussed in the main text is precisely the holonomy induced by this connection in parameter space.

\section{Quasi-Hermitian vs. Hermitian Uhlmann Connection and Phase under Similarity Transformations}
\label{subsec:qh_vs_h_uhlmann}

We have constructed parallel transport, the Uhlmann connection, and the induced geometric phase for mixed states within the quasi-Hermitian framework. This construction is based on the physical inner product defined by the positive-definite metric $\eta_+(\lambda)$ and takes the quasi-Hermitian parallel transport condition as the starting point for the geometric structure of mixed states.

In this framework, the introduction of a positive-definite metric implies that at each fixed parameter point, the quasi-Hermitian Hamiltonian can be mapped to a Hermitian operator under the Dirac inner product via an appropriate similarity transformation \cite{Hermiticity2003}. Thus, the quasi-Hermitian formulation is not independent of Hermitian quantum mechanics but rather constitutes an equivalent description reparameterized by the inner product structure. This naturally raises the question: under the correspondence established by the similarity transformation, does the Uhlmann connection (a local geometric object) defined in the quasi-Hermitian formulation and the induced Uhlmann phase (the global consequence of the holonomy) coincide with those in the corresponding Hermitian formulation? If not, how should the discrepancy be interpreted through the geometry of the purification space?

To concretely compare the geometric structures in the two formulations, we introduce a similarity transformation operator $S(\lambda)$ such that the quasi-Hermitian Hamiltonian $H(\lambda)$ and the Hermitian Hamiltonian $h(\lambda)$ satisfy
\begin{align}
h(\lambda) = S(\lambda)\,H(\lambda)\,S^{-1}(\lambda),\qquad h^\dagger(\lambda) = h(\lambda).
\label{eq:S_map_H_to_h_33}
\end{align}
Under the quasi-Hermitian conventions adopted in this paper, the metric operator can be taken as
\begin{align}
\eta_+(\lambda) = S^\dagger(\lambda)\,S(\lambda),
\label{eq:eta_S_relation_33}
\end{align}
so that under the similarity transformation $S$, an operator $A$ that is self-adjoint in the quasi-Hermitian sense ($A^{\ddagger}=A$) is mapped to a Hermitian operator in the Dirac sense ($(SAS^{-1})^\dagger = SAS^{-1}$).

From the perspective of Lie-group theory, the quasi-unitary group $U_{\eta_+(\lambda)}(N)$ defined by $U^{\ddagger}U=\mathbb I$ is isomorphic to the standard unitary group $U(N)$ via the similarity transformation $S(\lambda)$. Specifically, the conjugation map $\widetilde U = S U S^{-1}$ establishes a one-to-one correspondence between quasi-unitary and unitary operators. Hence, at each fixed parameter point, the two groups are identical as Lie groups. The distinction between the quasi-Hermitian and Hermitian formulations therefore does not arise from their algebraic structure, but rather from the way this symmetry group is fibered over the parameter space when $S(\lambda)$ (equivalently $\eta_+(\lambda)$) depends on the external parameters. This parameter-dependent fibration is precisely what gives rise to the additional geometric features explored in this work.

Under this similarity transformation, the biorthogonal eigenstates $\{|\Psi_n\rangle,\langle\Phi_n|\}$ of the quasi-Hermitian Hamiltonian are mapped to orthonormal eigenstates in the Hermitian formulation:
\begin{align}
|n(\lambda)\rangle := S(\lambda)\,|\Psi_n(\lambda)\rangle,\quad
\langle n(\lambda)| := \langle\Phi_n(\lambda)|\,S^{-1}(\lambda),
\label{eq:eigmap_33}
\end{align}
satisfying $h(\lambda)|n(\lambda)\rangle = E_n(\lambda)|n(\lambda)\rangle$ and $\langle n(\lambda)|m(\lambda)\rangle = \delta_{nm}$. Correspondingly, the density matrix in the Hermitian formulation is defined as
\begin{align}
\rho_h(\lambda) := S(\lambda)\,\rho(\lambda)\,S^{-1}(\lambda),
\label{eq:rho_map_33}
\end{align}
whose spectral decomposition is naturally given in the orthonormal basis $\{|n(\lambda)\rangle\}$.

In this Hermitian formulation, the geometric structure of mixed states can be constructed following the standard Uhlmann theory: the Uhlmann connection $\mathcal{A}_{\text{U}}^{h}$ corresponding to the path $\rho_h(\lambda)$ is uniquely determined by $\rho_h(\lambda)$ and takes the explicit form
\begin{align}
\mathcal{A}_{\text{U}}^{h} = -\sum_{m,n} |m\rangle \frac{\langle m|\,[\dif\sqrt{\rho_h},\sqrt{\rho_h}]\,|n\rangle}{p_m+p_n} \langle n|,
\label{eq:h_uhlmann_connection_33}
\end{align}
where $\{|n\rangle\}$ are the orthonormal eigenstates of $\rho_h$ and $p_n>0$ are the corresponding eigenvalues.

Comparing this with the Uhlmann connection obtained in the quasi-Hermitian formulation, Eq.~\eqref{eq:qh_A_explicit_final}, we see that the two expressions are structurally identical but are built on different inner product structures: the former acts on the density matrix path $\rho_h(\lambda)$ under the Dirac inner product, while the latter corresponds to the path $\rho(\lambda)$ under the physical inner product defined by the metric operator $\eta_+(\lambda)$.

If the similarity transformation $S$ is constant in parameter space (equivalently, if the metric operator $\eta_+$ does not vary with parameters), then $\rho_h$ and $\rho$ differ only by a fixed basis transformation. In this case, the Uhlmann connections in the two formulations satisfy the simple covariant relation
\begin{align}
\mathcal{A}_{\text{U}}^{\eta_+} = S^{-1} \mathcal{A}_{\text{U}}^{h} S,
\end{align}
and consequently the Uhlmann phases induced along a closed parameter path are identical after taking the trace.

When the similarity transformation $S(\lambda)$ is parameter-dependent, the correspondence between the quasi-Hermitian and Hermitian formulations changes fundamentally. In this case, $S(\lambda)$ is no longer merely a passive reparameterization; through the metric operator $\eta_+(\lambda)$, it introduces a parameter-dependent physical inner product structure. In the quasi-Hermitian formulation, the criterion for parallel transport is defined relative to a physical inner product that varies with parameters; in contrast, parallel transport in the Hermitian formulation is always based on the fixed Dirac inner product. This difference implies that the horizontal lifts of the path selected in the two formulations are generally not the same. Consequently, the similarity transformation induces an additional connection structure in the purification space, controlled by $\dif S\,S^{-1}$ (equivalently $\dif \eta_+$), such that the Uhlmann connections in the two formulations no longer satisfy a simple covariant relation.

In the quasi-Hermitian formulation, the Uhlmann connection $\mathcal A_{\text{U}}^{\eta_+}$ has been obtained in Eq.~\eqref{eq:qh_A_explicit_final}. In the Hermitian representation, the corresponding Uhlmann connection $\mathcal A_{\text{U}}^{h}$ associated with the density-matrix path $\rho_h(\lambda)$ is given by the standard expression Eq.~\eqref{eq:h_uhlmann_connection_33}. Using the similarity transformation relations $\rho_h = S\rho S^{-1}$ and $\sqrt{\rho_h} = S\sqrt{\rho}S^{-1}$, and applying the similarity transformation to Eq.~\eqref{eq:qh_A_explicit_final}, we obtain
\begin{align}
S \mathcal{A}_{\text{U}}^{\eta_+} S^{-1} = -\sum_{m,n} |m\rangle \frac{\langle m|\,S[\dif\sqrt{\rho},\sqrt{\rho}]S^{-1}\,|n\rangle}{p_m+p_n} \langle n|.
\label{eq:app_SAetaS}
\end{align}

The key difference arises from the differential of the square root operator. From $\sqrt{\rho_h} = S\sqrt{\rho}S^{-1}$, its differential is
\begin{align}
\dif\sqrt{\rho_h} = (\dif S)\sqrt{\rho}S^{-1} + S(\dif \sqrt{\rho})S^{-1} + S\sqrt{\rho}\dif(S^{-1}),
\end{align}
Introducing $K = \dif S\,S^{-1}$, the differential becomes
\begin{align}
\dif\sqrt{\rho_h} = K\sqrt{\rho_h} + S(\dif\sqrt{\rho})S^{-1} - \sqrt{\rho_h}K .
\end{align}
From this we compute the commutator
\begin{align}
[\dif\sqrt{\rho_h},\sqrt{\rho_h}] = S[\dif\sqrt{\rho},\sqrt{\rho}]S^{-1} + \Delta_S,
\label{eq:app_comm}
\end{align}
where
\begin{align}
\Delta_S = K\rho_h + \rho_h K - 2\sqrt{\rho_h}\,K\,\sqrt{\rho_h}.
\label{eq:app_DeltaS}
\end{align}

Substituting Eq.~\eqref{eq:app_comm} into the Hermitian Uhlmann connection \eqref{eq:h_uhlmann_connection_33} and comparing with Eq.~\eqref{eq:app_SAetaS} yields
\begin{align}
A_{\text{U}}^{h} = S \mathcal{A}_{\text{U}}^{\eta_+} S^{-1} - \mathcal A_S,
\end{align}
where the additional connection term is
\begin{align}
\mathcal A_S = \sum_{m,n} |m\rangle \frac{\langle m|\Delta_S|n\rangle}{p_m+p_n} \langle n|.
\label{eq:app_AS_operator}
\end{align}

In the eigenbasis of $\rho_h$, we have $\rho_h|n\rangle = p_n|n\rangle$ and $\sqrt{\rho_h}|n\rangle = \sqrt{p_n}|n\rangle$, so that
\begin{align}
\langle m|\Delta_S|n\rangle = (p_m + p_n - 2\sqrt{p_mp_n}) \langle m|K|n\rangle .
\end{align}
Inserting this into Eq.~\eqref{eq:app_AS_operator} gives
\begin{align}\label{eq:connection_difference_33}
\mathcal A_S = \sum_{m,n} \frac{(\sqrt{p_m}-\sqrt{p_n})^2}{p_m+p_n} \, |m\rangle\langle m| \dif S\,S^{-1} |n\rangle\langle n| .
\end{align}

This term does not originate from the density matrix path itself but reflects the variation of the physical inner product structure along the parameter space, projected onto the relevant Lie algebra directions via the Uhlmann parallel transport condition. When $\dif S = 0$ (equivalently $\dif \eta_+ = 0$), this additional term vanishes, restoring the strict equivalence between the quasi-Hermitian and Hermitian formulations at the level of the Uhlmann connection and phase.

It is important to emphasize that the Uhlmann connection considered here is a local path-dependent geometric object defined along the parameterized evolution, while the Uhlmann phase is the global quantity induced by the closed-loop parallel transport associated with this pathwise connection.

For a closed loop $\mathcal C$ in parameter space, the Uhlmann phase in the quasi-Hermitian formulation is defined by the holonomy of the connection $\mathcal{A}_{\text{U}}^{\eta_+}$, given by Eq.~\eqref{QHUP}, and remains invariant under quasi-unitary gauge transformations.
In the Hermitian formulation, the corresponding Uhlmann phase is determined by the holonomy of the connection $\mathcal{A}_{\text{U}}^{h}$.

From Eq.~\eqref{eq:connection_difference_33}, the local difference between the two connections is governed by the metric-induced term $\mathcal A_S$. This term modifies the effective connection in the Hermitian representation and thereby affects the closed-loop parallel transport, as described by the holonomy
\begin{align}
U_h(C)
=
\mathcal P
\exp\left[
-\oint_C
\left(
S A_U^{\eta_+}S^{-1}-A_S
\right)
\right].
\end{align}
Thus, in general, the influence of $\mathcal A_S$ is incorporated through the full path-ordered evolution. In the Abelian case, or whenever the relevant connection components commute along the loop, the holonomy factorizes as
\begin{align}
U_h(C)
=
\exp\left[
-\oint_C S A_U^{\eta_+}S^{-1}
\right]
\exp\left[
\oint_C A_S
\right].
\end{align}
In this restricted case, after projection onto the corresponding Uhlmann phase, the metric-induced factor defines an effective scalar loop contribution $\Gamma_S$, giving
\begin{align}
\theta_{\text{U}}^{h}
=
\theta_{\text{U}}^{\eta_+}
+
\oint_{\mathcal C}
\Gamma_S .
\end{align}

When the metric reduces to the identity operator, the additional term $\mathcal A_S$ vanishes, and the above structure continuously reduces to the standard Uhlmann theory in Hermitian quantum mechanics.

In summary, this appendix demonstrates that while the similarity transformation ensures consistency between the quasi-Hermitian and Hermitian formulations at the level of spectral and probabilistic structures, at the level of the mixed-state geometric structure, a parameter-dependent similarity transformation introduces an additional connection term that modifies the parallel transport rule, leading to nontrivial deviations in the Uhlmann connection and consequently in the Uhlmann phase.

\section{Derivation of the Uhlmann Holonomy on an Equatorial Loop}
\label{app:equator_holonomy}
Here we provide the analytic derivation of the Uhlmann holonomy for an equatorial closed loop given in Eq.~\eqref{Uhlholonomy} of the main text.
The holonomy along the equatorial path $\phi\in[0,2\pi\Omega]$ is defined as
\begin{align}
U(2\pi\Omega)
=
\mathcal{P}\exp\!\left(-
\int_{0}^{2\pi\Omega} A_\phi(\phi)\,\dif\phi
\right),
\end{align}
where $A_{\text{U}}^{\eta_+}=A_\phi(\phi)\,\dif\phi$ and
\begin{align}
A_\phi(\phi)
=
\kappa
\left[
\frac{2a^2}{\Delta^2}\,\mi\,\sigma_z
-
\frac{2ab}{\Delta^2}
\me^{-\mi\phi\sigma_z/2}\sigma_x \me^{\mi\phi\sigma_z/2}
\right].
\label{eq:Aphi_app}
\end{align}
The path-ordered exponential is equivalent to solving the initial value problem
\begin{align}
\frac{\dif}{\dif\phi}U(\phi)
=
-A_\phi(\phi)U(\phi),
\quad
U(0)=\mathbb{I}.
\label{eq:ODE_app}
\end{align}

To remove the explicit $\phi$ dependence of $A_\phi(\phi)$, we introduce the rotation operator about the $z$-axis:
\begin{align}
R(\phi)
=
\exp\!\left(-\frac{\mi\phi}{2}\sigma_z\right),
\label{eq:R_def_app}
\end{align}
and perform a change of variables to a rotating frame:
\begin{align}
U(\phi)
=
R(\phi)\,\widetilde U(\phi).
\label{eq:gauge_transform_app}
\end{align}
Substituting Eq.~\eqref{eq:gauge_transform_app} into Eq.~\eqref{eq:ODE_app} and multiplying on the left by $R^{-1}(\phi)$ yields
\begin{align}
\frac{\dif}{\dif\phi}\widetilde U(\phi)
&=
\widetilde A_\phi\,\widetilde U(\phi), \notag \\
\widetilde A_\phi
:&=
-R^{-1}A_\phi R
-
R^{-1}\frac{\dif R}{\dif\phi}.
\label{eq:Atilde_def_app}
\end{align}

Using $R^{-1}(\phi)\sigma_z R(\phi)=\sigma_z$, we obtain
\begin{align}
R^{-1}(\phi)A_\phi(\phi)R(\phi)
=
\kappa
\left[
\frac{2a^2}{\Delta^2}\,\mi\,\sigma_z
-
\frac{2ab}{\Delta^2}\sigma_x
\right].
\label{eq:RAR_app}
\end{align}
Furthermore, from Eq.~\eqref{eq:R_def_app},
\begin{align}
\frac{\dif R}{\dif\phi}
=
-\frac{\mi}{2}\sigma_z R(\phi),
\qquad
R^{-1}\frac{\dif R}{\dif\phi}
=
-\frac{\mi}{2}\sigma_z .
\label{eq:RinvRprime_app}
\end{align}
Substituting into Eq.~\eqref{eq:Atilde_def_app} gives a $\phi$-independent constant generator
\begin{align}
\widetilde A_\phi
=
\mi\left(
\frac{1}{2}-\frac{2a^2}{\Delta^2}\kappa
\right)\sigma_z
+
\frac{2ab}{\Delta^2}\kappa\,\sigma_x .
\label{eq:Atilde_final_app}
\end{align}
Therefore,
\begin{align}
\widetilde U(\phi)
=
\exp\!\left(\phi\,\widetilde A_\phi\right).
\end{align}
and consequently
\begin{align}
U(2\pi\Omega)
=
R(2\pi\Omega)
\exp\!\left(2\pi\Omega\,\widetilde A_\phi\right).
\end{align}
Noting that
\begin{align}
R(2\pi\Omega)
=
\exp(-i\pi\Omega\sigma_z)
=
(-1)^\Omega\mathbb{I},
\end{align}
we finally obtain
\begin{align}
U(2\pi\Omega)= (-1)^{\Omega}\me^{2\pi\Omega\Big[\mi\Big(\frac{1}{2}-\frac{2a^2\kappa}{\Delta^2}\Big)\sigma_{z}+\frac{2ab\kappa}{\Delta^2}\sigma_x\Big]},
\end{align}

\section{Equivalence between the Quasi-Hermitian Hilbert--Schmidt Product and the Loschmidt Amplitude}
\label{app:purification_proof}
Here we prove that, under the vectorization (purification) map, the quasi-Hermitian Hilbert--Schmidt inner product between two amplitudes is equivalent to the overlap (Loschmidt amplitude/fidelity) between the corresponding purified states in the enlarged space:
\begin{align}
\mathrm{Tr}_{\eta_+}\!\left(W_1^{\ddagger} W_2\right)
=
\langle W_1 | W_2 \rangle_{\eta_+ \otimes \eta_+^{-1}} .
\label{eq:HS_Loschmidt_equiv_app}
\end{align}

We start from the spectral-form purifications
\begin{align}
W=\sum_n \sqrt{p_n}\,|\psi_n\rangle\langle\phi_n|\,\mathcal U,
\end{align}
and define the corresponding purified states
\begin{align}
|W\rangle
=
\sum_n \sqrt{p_n}\,
|\psi_n\rangle_S
\otimes
\mathcal U^{T}|\phi_n\rangle_A .
\label{eq:vec_def_app}
\end{align}
The metric-weighted inner product in the enlarged space is
\begin{align}
\langle W_1|W_2\rangle_{\eta_+ \otimes \eta_+^{-1}}
:=
\langle W_1|\big(\eta_+\otimes \eta_+^{-1}\big)|W_2\rangle .
\end{align}

Substituting Eq.~\eqref{eq:vec_def_app} gives
\begin{align}
&\langle W_1|W_2 \rangle_{\eta_+ \otimes \eta_+^{-1}} \notag \\
&=\sum_{mn} \sqrt{p_n}\sqrt{p_m}
\langle\psi_n|\eta_+|\psi_m\rangle
\langle \phi_m|\mathcal{U}_2 \eta_+^{-1} \mathcal{U}_1^{\dag}| \phi_n\rangle \notag \\
&=\sum_{mn} p_n\delta_{mn}
\langle \phi_m|\mathcal{U}_2 \eta_+^{-1} \mathcal{U}_1^{\dag}\eta_+| \psi_n\rangle \notag \\
&=\sum_{n} p_n
\langle \phi_n|\mathcal{U}_2 \mathcal{U}_1^{\ddag}| \psi_n\rangle \notag \\
&=\mathrm{Tr}_{\eta_+}(\rho\mathcal{U}_2\mathcal{U}^{\ddag}_1).
\label{eq:step1_app}
\end{align}
It is important to note that, in the above derivation, the factor $\mathcal{U}^T|\phi_n\rangle$ in the purified state is to be treated as a bra.
On the other hand, a direct evaluation of the quasi-Hermitian Hilbert--Schmidt product yields
\begin{align}
&\mathrm{Tr}_{\eta_+}(W^{\ddag}_1 W_2) \notag \\
&=\mathrm{Tr}_{\eta_+}(\sum_n \mathcal{U}^{\ddag}_1\sqrt{p_n}\eta^{-1}|\phi_n\rangle \langle \psi_n|\eta\sum_m \sqrt{p_m}|\psi_m\rangle \langle \phi_m|\mathcal{U}_2) \notag \\
&=\mathrm{Tr}_{\eta_+}(\sum_n \mathcal{U}^{\ddag}_1\sqrt{p_n}|\psi_n\rangle \langle \phi_n|\sum_m \sqrt{p_m}|\psi_m\rangle \langle \phi_m|\mathcal{U}_2) \notag \\
&=\mathrm{Tr}_{\eta_+}(\sum_{m,n} \mathcal{U}^{\ddag}_1\sqrt{p_n}\sqrt{p_m}|\psi_n\rangle \delta_{mn} \langle \phi_m|\mathcal{U}_2) \notag \\
&=\mathrm{Tr}_{\eta_+}(\sum_n p_n|\psi_n\rangle \langle \phi_n|\mathcal{U}_2\mathcal{U}^{\ddag}_1) \notag \\
&=\mathrm{Tr}_{\eta_+}(\rho\mathcal{U}_2\mathcal{U}^{\ddag}_1).
\end{align}
Therefore, the geometric amplitude can be viewed as the Loschmidt (fidelity) amplitude between the initial and final purifications.
\bibliography{Review}

\begin{thebibliography}{67}%
\makeatletter
\providecommand \@ifxundefined [1]{%
 \@ifx{#1\undefined}
}%
\providecommand \@ifnum [1]{%
 \ifnum #1\expandafter \@firstoftwo
 \else \expandafter \@secondoftwo
 \fi
}%
\providecommand \@ifx [1]{%
 \ifx #1\expandafter \@firstoftwo
 \else \expandafter \@secondoftwo
 \fi
}%
\providecommand \natexlab [1]{#1}%
\providecommand \enquote  [1]{``#1''}%
\providecommand \bibnamefont  [1]{#1}%
\providecommand \bibfnamefont [1]{#1}%
\providecommand \citenamefont [1]{#1}%
\providecommand \href@noop [0]{\@secondoftwo}%
\providecommand \href [0]{\begingroup \@sanitize@url \@href}%
\providecommand \@href[1]{\@@startlink{#1}\@@href}%
\providecommand \@@href[1]{\endgroup#1\@@endlink}%
\providecommand \@sanitize@url [0]{\catcode `\\12\catcode `\$12\catcode
  `\&12\catcode `\#12\catcode `\^12\catcode `\_12\catcode `\%12\relax}%
\providecommand \@@startlink[1]{}%
\providecommand \@@endlink[0]{}%
\providecommand \url  [0]{\begingroup\@sanitize@url \@url }%
\providecommand \@url [1]{\endgroup\@href {#1}{\urlprefix }}%
\providecommand \urlprefix  [0]{URL }%
\providecommand \Eprint [0]{\href }%
\providecommand \doibase [0]{https://doi.org/}%
\providecommand \selectlanguage [0]{\@gobble}%
\providecommand \bibinfo  [0]{\@secondoftwo}%
\providecommand \bibfield  [0]{\@secondoftwo}%
\providecommand \translation [1]{[#1]}%
\providecommand \BibitemOpen [0]{}%
\providecommand \bibitemStop [0]{}%
\providecommand \bibitemNoStop [0]{.\EOS\space}%
\providecommand \EOS [0]{\spacefactor3000\relax}%
\providecommand \BibitemShut  [1]{\csname bibitem#1\endcsname}%
\let\auto@bib@innerbib\@empty
%</preamble>
\bibitem [{\citenamefont {Pancharatnam}(1956)}]{Pancharatnam1956}%
  \BibitemOpen
  \bibfield  {author} {\bibinfo {author} {\bibfnamefont {S.}~\bibnamefont
  {Pancharatnam}},\ }\bibfield  {title} {\bibinfo {title} {Generalized theory
  of interference, and its applications},\ }\href
  {https://doi.org/10.1007/BF03046050} {\bibfield  {journal} {\bibinfo
  {journal} {Proceedings of the Indian Academy of Sciences - Section A}\
  }\textbf {\bibinfo {volume} {44}},\ \bibinfo {pages} {247} (\bibinfo {year}
  {1956})}\BibitemShut {NoStop}%
\bibitem [{\citenamefont {Berry}(1984)}]{Berry1984}%
  \BibitemOpen
  \bibfield  {author} {\bibinfo {author} {\bibfnamefont {M.~V.}\ \bibnamefont
  {Berry}},\ }\bibfield  {title} {\bibinfo {title} {Quantal phase factors
  accompanying adiabatic changes},\ }\href
  {https://doi.org/10.1098/rspa.1984.0023} {\bibfield  {journal} {\bibinfo
  {journal} {Proceedings of the Royal Society of London A}\ }\textbf {\bibinfo
  {volume} {392}},\ \bibinfo {pages} {45} (\bibinfo {year} {1984})}\BibitemShut
  {NoStop}%
\bibitem [{\citenamefont {Sj\"oqvist}\ \emph {et~al.}(2000)\citenamefont
  {Sj\"oqvist}, \citenamefont {Pati}, \citenamefont {Ekert}, \citenamefont
  {Anandan}, \citenamefont {Ericsson}, \citenamefont {Oi},\ and\ \citenamefont
  {Vedral}}]{PhysRevLett.85.2845}%
  \BibitemOpen
  \bibfield  {author} {\bibinfo {author} {\bibfnamefont {E.}~\bibnamefont
  {Sj\"oqvist}}, \bibinfo {author} {\bibfnamefont {A.~K.}\ \bibnamefont
  {Pati}}, \bibinfo {author} {\bibfnamefont {A.}~\bibnamefont {Ekert}},
  \bibinfo {author} {\bibfnamefont {J.~S.}\ \bibnamefont {Anandan}}, \bibinfo
  {author} {\bibfnamefont {M.}~\bibnamefont {Ericsson}}, \bibinfo {author}
  {\bibfnamefont {D.~K.~L.}\ \bibnamefont {Oi}},\ and\ \bibinfo {author}
  {\bibfnamefont {V.}~\bibnamefont {Vedral}},\ }\bibfield  {title} {\bibinfo
  {title} {Geometric phases for mixed states in interferometry},\ }\href
  {https://doi.org/10.1103/PhysRevLett.85.2845} {\bibfield  {journal} {\bibinfo
   {journal} {Phys. Rev. Lett.}\ }\textbf {\bibinfo {volume} {85}},\ \bibinfo
  {pages} {2845} (\bibinfo {year} {2000})}\BibitemShut {NoStop}%
\bibitem [{\citenamefont {Ericsson}\ \emph {et~al.}(2003)\citenamefont
  {Ericsson}, \citenamefont {Sj\"oqvist}, \citenamefont {Br\"annlund},
  \citenamefont {Oi},\ and\ \citenamefont {Pati}}]{PhysRevA.67.020101}%
  \BibitemOpen
  \bibfield  {author} {\bibinfo {author} {\bibfnamefont {M.}~\bibnamefont
  {Ericsson}}, \bibinfo {author} {\bibfnamefont {E.}~\bibnamefont
  {Sj\"oqvist}}, \bibinfo {author} {\bibfnamefont {J.}~\bibnamefont
  {Br\"annlund}}, \bibinfo {author} {\bibfnamefont {D.~K.~L.}\ \bibnamefont
  {Oi}},\ and\ \bibinfo {author} {\bibfnamefont {A.~K.}\ \bibnamefont {Pati}},\
  }\bibfield  {title} {\bibinfo {title} {Generalization of the geometric phase
  to completely positive maps},\ }\href
  {https://doi.org/10.1103/PhysRevA.67.020101} {\bibfield  {journal} {\bibinfo
  {journal} {Phys. Rev. A}\ }\textbf {\bibinfo {volume} {67}},\ \bibinfo
  {pages} {020101} (\bibinfo {year} {2003})}\BibitemShut {NoStop}%
\bibitem [{\citenamefont {Carollo}\ \emph {et~al.}(2003)\citenamefont
  {Carollo}, \citenamefont {Fuentes-Guridi}, \citenamefont {Santos},\ and\
  \citenamefont {Vedral}}]{PhysRevLett.90.160402}%
  \BibitemOpen
  \bibfield  {author} {\bibinfo {author} {\bibfnamefont {A.}~\bibnamefont
  {Carollo}}, \bibinfo {author} {\bibfnamefont {I.}~\bibnamefont
  {Fuentes-Guridi}}, \bibinfo {author} {\bibfnamefont {M.~F. m.~c.}\
  \bibnamefont {Santos}},\ and\ \bibinfo {author} {\bibfnamefont
  {V.}~\bibnamefont {Vedral}},\ }\bibfield  {title} {\bibinfo {title}
  {Geometric phase in open systems},\ }\href
  {https://doi.org/10.1103/PhysRevLett.90.160402} {\bibfield  {journal}
  {\bibinfo  {journal} {Phys. Rev. Lett.}\ }\textbf {\bibinfo {volume} {90}},\
  \bibinfo {pages} {160402} (\bibinfo {year} {2003})}\BibitemShut {NoStop}%
\bibitem [{\citenamefont {Faria}\ \emph {et~al.}(2003)\citenamefont {Faria},
  \citenamefont {Piza},\ and\ \citenamefont {Nemes}}]{Faria_2003}%
  \BibitemOpen
  \bibfield  {author} {\bibinfo {author} {\bibfnamefont {J.~G. P.~d.}\
  \bibnamefont {Faria}}, \bibinfo {author} {\bibfnamefont {A.~F. R. d.~T.}\
  \bibnamefont {Piza}},\ and\ \bibinfo {author} {\bibfnamefont {M.~C.}\
  \bibnamefont {Nemes}},\ }\bibfield  {title} {\bibinfo {title} {Phases of
  quantum states in completely positive non-unitary evolution},\ }\href
  {https://doi.org/10.1209/epl/i2003-00440-4} {\bibfield  {journal} {\bibinfo
  {journal} {Europhysics Letters (EPL)}\ }\textbf {\bibinfo {volume} {62}},\
  \bibinfo {pages} {782–788} (\bibinfo {year} {2003})}\BibitemShut {NoStop}%
\bibitem [{\citenamefont {Du}\ \emph {et~al.}(2003)\citenamefont {Du},
  \citenamefont {Zou}, \citenamefont {Shi}, \citenamefont {Kwek}, \citenamefont
  {Pan}, \citenamefont {Oh}, \citenamefont {Ekert}, \citenamefont {Oi},\ and\
  \citenamefont {Ericsson}}]{PhysRevLett.91.100403}%
  \BibitemOpen
  \bibfield  {author} {\bibinfo {author} {\bibfnamefont {J.}~\bibnamefont
  {Du}}, \bibinfo {author} {\bibfnamefont {P.}~\bibnamefont {Zou}}, \bibinfo
  {author} {\bibfnamefont {M.}~\bibnamefont {Shi}}, \bibinfo {author}
  {\bibfnamefont {L.~C.}\ \bibnamefont {Kwek}}, \bibinfo {author}
  {\bibfnamefont {J.-W.}\ \bibnamefont {Pan}}, \bibinfo {author} {\bibfnamefont
  {C.~H.}\ \bibnamefont {Oh}}, \bibinfo {author} {\bibfnamefont
  {A.}~\bibnamefont {Ekert}}, \bibinfo {author} {\bibfnamefont {D.~K.~L.}\
  \bibnamefont {Oi}},\ and\ \bibinfo {author} {\bibfnamefont {M.}~\bibnamefont
  {Ericsson}},\ }\bibfield  {title} {\bibinfo {title} {Observation of geometric
  phases for mixed states using nmr interferometry},\ }\href
  {https://doi.org/10.1103/PhysRevLett.91.100403} {\bibfield  {journal}
  {\bibinfo  {journal} {Phys. Rev. Lett.}\ }\textbf {\bibinfo {volume} {91}},\
  \bibinfo {pages} {100403} (\bibinfo {year} {2003})}\BibitemShut {NoStop}%
\bibitem [{\citenamefont {Klepp}\ \emph {et~al.}(2008)\citenamefont {Klepp},
  \citenamefont {Sponar}, \citenamefont {Filipp}, \citenamefont {Lettner},
  \citenamefont {Badurek},\ and\ \citenamefont
  {Hasegawa}}]{PhysRevLett.101.150404}%
  \BibitemOpen
  \bibfield  {author} {\bibinfo {author} {\bibfnamefont {J.}~\bibnamefont
  {Klepp}}, \bibinfo {author} {\bibfnamefont {S.}~\bibnamefont {Sponar}},
  \bibinfo {author} {\bibfnamefont {S.}~\bibnamefont {Filipp}}, \bibinfo
  {author} {\bibfnamefont {M.}~\bibnamefont {Lettner}}, \bibinfo {author}
  {\bibfnamefont {G.}~\bibnamefont {Badurek}},\ and\ \bibinfo {author}
  {\bibfnamefont {Y.}~\bibnamefont {Hasegawa}},\ }\bibfield  {title} {\bibinfo
  {title} {Observation of nonadditive mixed-state phases with polarized
  neutrons},\ }\href {https://doi.org/10.1103/PhysRevLett.101.150404}
  {\bibfield  {journal} {\bibinfo  {journal} {Phys. Rev. Lett.}\ }\textbf
  {\bibinfo {volume} {101}},\ \bibinfo {pages} {150404} (\bibinfo {year}
  {2008})}\BibitemShut {NoStop}%
\bibitem [{\citenamefont {Ericsson}\ \emph {et~al.}(2005)\citenamefont
  {Ericsson}, \citenamefont {Achilles}, \citenamefont {Barreiro}, \citenamefont
  {Branning}, \citenamefont {Peters},\ and\ \citenamefont
  {Kwiat}}]{PhysRevLett.94.050401}%
  \BibitemOpen
  \bibfield  {author} {\bibinfo {author} {\bibfnamefont {M.}~\bibnamefont
  {Ericsson}}, \bibinfo {author} {\bibfnamefont {D.}~\bibnamefont {Achilles}},
  \bibinfo {author} {\bibfnamefont {J.~T.}\ \bibnamefont {Barreiro}}, \bibinfo
  {author} {\bibfnamefont {D.}~\bibnamefont {Branning}}, \bibinfo {author}
  {\bibfnamefont {N.~A.}\ \bibnamefont {Peters}},\ and\ \bibinfo {author}
  {\bibfnamefont {P.~G.}\ \bibnamefont {Kwiat}},\ }\bibfield  {title} {\bibinfo
  {title} {Measurement of geometric phase for mixed states using single photon
  interferometry},\ }\href {https://doi.org/10.1103/PhysRevLett.94.050401}
  {\bibfield  {journal} {\bibinfo  {journal} {Phys. Rev. Lett.}\ }\textbf
  {\bibinfo {volume} {94}},\ \bibinfo {pages} {050401} (\bibinfo {year}
  {2005})}\BibitemShut {NoStop}%
\bibitem [{\citenamefont {Uhlmann}(1986)}]{Uhlmann1986}%
  \BibitemOpen
  \bibfield  {author} {\bibinfo {author} {\bibfnamefont {A.}~\bibnamefont
  {Uhlmann}},\ }\bibfield  {title} {\bibinfo {title} {Parallel transport and
  “quantum holonomy” along density operators},\ }\href
  {https://doi.org/10.1016/0034-4877(86)90055-8} {\bibfield  {journal}
  {\bibinfo  {journal} {Rep. Math. Phys.}\ }\textbf {\bibinfo {volume} {24}},\
  \bibinfo {pages} {229} (\bibinfo {year} {1986})}\BibitemShut {NoStop}%
\bibitem [{\citenamefont {Uhlmann}(1989)}]{Uhlmann1989}%
  \BibitemOpen
  \bibfield  {author} {\bibinfo {author} {\bibfnamefont {A.}~\bibnamefont
  {Uhlmann}},\ }\bibfield  {title} {\bibinfo {title} {On berry phases along
  mixtures of states},\ }\href {https://doi.org/10.1002/andp.19895010108}
  {\bibfield  {journal} {\bibinfo  {journal} {Ann. Phys. (Berlin)}\ }\textbf
  {\bibinfo {volume} {501}},\ \bibinfo {pages} {63} (\bibinfo {year}
  {1989})}\BibitemShut {NoStop}%
\bibitem [{\citenamefont {Uhlmann}(1991)}]{Uhlmann1991}%
  \BibitemOpen
  \bibfield  {author} {\bibinfo {author} {\bibfnamefont {A.}~\bibnamefont
  {Uhlmann}},\ }\bibfield  {title} {\bibinfo {title} {A gauge field governing
  parallel transport along mixed states},\ }\href
  {https://doi.org/10.1007/BF00420373} {\bibfield  {journal} {\bibinfo
  {journal} {Lett. Math. Phys.}\ }\textbf {\bibinfo {volume} {21}},\ \bibinfo
  {pages} {229} (\bibinfo {year} {1991})}\BibitemShut {NoStop}%
\bibitem [{\citenamefont {Uhlmann}(1995)}]{UHLMANN1995461}%
  \BibitemOpen
  \bibfield  {author} {\bibinfo {author} {\bibfnamefont {A.}~\bibnamefont
  {Uhlmann}},\ }\bibfield  {title} {\bibinfo {title} {Geometric phases and
  related structures},\ }\href@noop {} {\bibfield  {journal} {\bibinfo
  {journal} {Reports on Mathematical Physics}\ }\textbf {\bibinfo {volume}
  {36}},\ \bibinfo {pages} {461} (\bibinfo {year} {1995})}\BibitemShut
  {NoStop}%
\bibitem [{\citenamefont {Viyuela}\ \emph
  {et~al.}(2014{\natexlab{a}})\citenamefont {Viyuela}, \citenamefont {Rivas},\
  and\ \citenamefont {Martin-Delgado}}]{Viyuela14}%
  \BibitemOpen
  \bibfield  {author} {\bibinfo {author} {\bibfnamefont {O.}~\bibnamefont
  {Viyuela}}, \bibinfo {author} {\bibfnamefont {A.}~\bibnamefont {Rivas}},\
  and\ \bibinfo {author} {\bibfnamefont {M.~A.}\ \bibnamefont
  {Martin-Delgado}},\ }\bibfield  {title} {\bibinfo {title} {Uhlmann phase as a
  topological measure for one-dimensional fermion systems},\ }\href@noop {}
  {\bibfield  {journal} {\bibinfo  {journal} {Phys. Rev. Lett.}\ }\textbf
  {\bibinfo {volume} {112}},\ \bibinfo {pages} {130401} (\bibinfo {year}
  {2014}{\natexlab{a}})}\BibitemShut {NoStop}%
\bibitem [{\citenamefont {Viyuela}\ \emph
  {et~al.}(2014{\natexlab{b}})\citenamefont {Viyuela}, \citenamefont {Rivas},\
  and\ \citenamefont {Martin-Delgado}}]{ViyuelaPRL14-2}%
  \BibitemOpen
  \bibfield  {author} {\bibinfo {author} {\bibfnamefont {O.}~\bibnamefont
  {Viyuela}}, \bibinfo {author} {\bibfnamefont {A.}~\bibnamefont {Rivas}},\
  and\ \bibinfo {author} {\bibfnamefont {M.~A.}\ \bibnamefont
  {Martin-Delgado}},\ }\bibfield  {title} {\bibinfo {title} {Two-dimensional
  density-matrix topological fermionic phases: Topological uhlmann numbers},\
  }\href@noop {} {\bibfield  {journal} {\bibinfo  {journal} {Phys. Rev. Lett.}\
  }\textbf {\bibinfo {volume} {113}},\ \bibinfo {pages} {076408} (\bibinfo
  {year} {2014}{\natexlab{b}})}\BibitemShut {NoStop}%
\bibitem [{\citenamefont {Guo}\ \emph {et~al.}(2020)\citenamefont {Guo},
  \citenamefont {Hou}, \citenamefont {He},\ and\ \citenamefont
  {Chien}}]{ourPRB20}%
  \BibitemOpen
  \bibfield  {author} {\bibinfo {author} {\bibfnamefont {H.}~\bibnamefont
  {Guo}}, \bibinfo {author} {\bibfnamefont {X.-Y.}\ \bibnamefont {Hou}},
  \bibinfo {author} {\bibfnamefont {Y.}~\bibnamefont {He}},\ and\ \bibinfo
  {author} {\bibfnamefont {C.-C.}\ \bibnamefont {Chien}},\ }\bibfield  {title}
  {\bibinfo {title} {Dynamic process and uhlmann process: Incompatibility and
  dynamic phase of mixed quantum states},\ }\href@noop {} {\bibfield  {journal}
  {\bibinfo  {journal} {Phys. Rev. B}\ }\textbf {\bibinfo {volume} {101}},\
  \bibinfo {pages} {104310} (\bibinfo {year} {2020})}\BibitemShut {NoStop}%
\bibitem [{\citenamefont {Hou}\ \emph {et~al.}(2020)\citenamefont {Hou},
  \citenamefont {Gao}, \citenamefont {Guo}, \citenamefont {He}, \citenamefont
  {Liu},\ and\ \citenamefont {Chien}}]{OurPRB20b}%
  \BibitemOpen
  \bibfield  {author} {\bibinfo {author} {\bibfnamefont {X.-Y.}\ \bibnamefont
  {Hou}}, \bibinfo {author} {\bibfnamefont {Q.-C.}\ \bibnamefont {Gao}},
  \bibinfo {author} {\bibfnamefont {H.}~\bibnamefont {Guo}}, \bibinfo {author}
  {\bibfnamefont {Y.}~\bibnamefont {He}}, \bibinfo {author} {\bibfnamefont
  {T.}~\bibnamefont {Liu}},\ and\ \bibinfo {author} {\bibfnamefont {C.-C.}\
  \bibnamefont {Chien}},\ }\bibfield  {title} {\bibinfo {title} {Ubiquity of
  zeros of the loschmidt amplitude for mixed states in different physical
  processes and its implication},\ }\href@noop {} {\bibfield  {journal}
  {\bibinfo  {journal} {Phys. Rev. B}\ }\textbf {\bibinfo {volume} {102}},\
  \bibinfo {pages} {104305} (\bibinfo {year} {2020})}\BibitemShut {NoStop}%
\bibitem [{\citenamefont {Morachis~Galindo}\ \emph {et~al.}(2021)\citenamefont
  {Morachis~Galindo}, \citenamefont {Rojas},\ and\ \citenamefont
  {Maytorena}}]{Galindo21}%
  \BibitemOpen
  \bibfield  {author} {\bibinfo {author} {\bibfnamefont {D.}~\bibnamefont
  {Morachis~Galindo}}, \bibinfo {author} {\bibfnamefont {F.}~\bibnamefont
  {Rojas}},\ and\ \bibinfo {author} {\bibfnamefont {J.~A.}\ \bibnamefont
  {Maytorena}},\ }\bibfield  {title} {\bibinfo {title} {Topological uhlmann
  phase transitions for a spin-$j$ particle in a magnetic field},\ }\href@noop
  {} {\bibfield  {journal} {\bibinfo  {journal} {Phys. Rev. A}\ }\textbf
  {\bibinfo {volume} {103}},\ \bibinfo {pages} {042221} (\bibinfo {year}
  {2021})}\BibitemShut {NoStop}%
\bibitem [{\citenamefont {Zhang}\ \emph {et~al.}(2021)\citenamefont {Zhang},
  \citenamefont {Pi}, \citenamefont {He},\ and\ \citenamefont
  {Chien}}]{Zhang21}%
  \BibitemOpen
  \bibfield  {author} {\bibinfo {author} {\bibfnamefont {Y.}~\bibnamefont
  {Zhang}}, \bibinfo {author} {\bibfnamefont {A.}~\bibnamefont {Pi}}, \bibinfo
  {author} {\bibfnamefont {Y.}~\bibnamefont {He}},\ and\ \bibinfo {author}
  {\bibfnamefont {C.-C.}\ \bibnamefont {Chien}},\ }\bibfield  {title} {\bibinfo
  {title} {Comparison of finite-temperature topological indicators based on
  uhlmann connection},\ }\href@noop {} {\bibfield  {journal} {\bibinfo
  {journal} {Phys. Rev. B}\ }\textbf {\bibinfo {volume} {104}},\ \bibinfo
  {pages} {165417} (\bibinfo {year} {2021})}\BibitemShut {NoStop}%
\bibitem [{\citenamefont {Hou}\ \emph {et~al.}(2021)\citenamefont {Hou},
  \citenamefont {Guo},\ and\ \citenamefont {Chien}}]{PhysRevA.104.023303}%
  \BibitemOpen
  \bibfield  {author} {\bibinfo {author} {\bibfnamefont {X.-Y.}\ \bibnamefont
  {Hou}}, \bibinfo {author} {\bibfnamefont {H.}~\bibnamefont {Guo}},\ and\
  \bibinfo {author} {\bibfnamefont {C.-C.}\ \bibnamefont {Chien}},\ }\bibfield
  {title} {\bibinfo {title} {Finite-temperature topological phase transitions
  of spin-$j$ systems in uhlmann processes: General formalism and experimental
  protocols},\ }\href@noop {} {\bibfield  {journal} {\bibinfo  {journal} {Phys.
  Rev. A}\ }\textbf {\bibinfo {volume} {104}},\ \bibinfo {pages} {023303}
  (\bibinfo {year} {2021})}\BibitemShut {NoStop}%
\bibitem [{\citenamefont {Hou}\ \emph {et~al.}(2023)\citenamefont {Hou},
  \citenamefont {Wang}, \citenamefont {Zhou}, \citenamefont {Guo},\ and\
  \citenamefont {Chien}}]{Hou2023}%
  \BibitemOpen
  \bibfield  {author} {\bibinfo {author} {\bibfnamefont {X.-Y.}\ \bibnamefont
  {Hou}}, \bibinfo {author} {\bibfnamefont {X.}~\bibnamefont {Wang}}, \bibinfo
  {author} {\bibfnamefont {Z.}~\bibnamefont {Zhou}}, \bibinfo {author}
  {\bibfnamefont {H.}~\bibnamefont {Guo}},\ and\ \bibinfo {author}
  {\bibfnamefont {C.-C.}\ \bibnamefont {Chien}},\ }\bibfield  {title} {\bibinfo
  {title} {Geometric phases of mixed quantum states: A comparative study of
  interferometric and uhlmann phases},\ }\href
  {https://doi.org/10.1103/PhysRevB.107.165415} {\bibfield  {journal} {\bibinfo
   {journal} {Phys. Rev. B}\ }\textbf {\bibinfo {volume} {107}},\ \bibinfo
  {pages} {165415} (\bibinfo {year} {2023})}\BibitemShut {NoStop}%
\bibitem [{\citenamefont {Huang}\ and\ \citenamefont
  {Arovas}(2014)}]{PhysRevLett.113.076407}%
  \BibitemOpen
  \bibfield  {author} {\bibinfo {author} {\bibfnamefont {Z.}~\bibnamefont
  {Huang}}\ and\ \bibinfo {author} {\bibfnamefont {D.~P.}\ \bibnamefont
  {Arovas}},\ }\bibfield  {title} {\bibinfo {title} {Topological indices for
  open and thermal systems via uhlmann's phase},\ }\href@noop {} {\bibfield
  {journal} {\bibinfo  {journal} {Phys. Rev. Lett.}\ }\textbf {\bibinfo
  {volume} {113}},\ \bibinfo {pages} {076407} (\bibinfo {year}
  {2014})}\BibitemShut {NoStop}%
\bibitem [{\citenamefont {Kiselev}\ and\ \citenamefont
  {Kesaev}(2018)}]{PhysRevA.98.033816}%
  \BibitemOpen
  \bibfield  {author} {\bibinfo {author} {\bibfnamefont {A.~D.}\ \bibnamefont
  {Kiselev}}\ and\ \bibinfo {author} {\bibfnamefont {V.~V.}\ \bibnamefont
  {Kesaev}},\ }\bibfield  {title} {\bibinfo {title} {Interferometric and
  uhlmann phases of mixed polarization states},\ }\href@noop {} {\bibfield
  {journal} {\bibinfo  {journal} {Phys. Rev. A}\ }\textbf {\bibinfo {volume}
  {98}},\ \bibinfo {pages} {033816} (\bibinfo {year} {2018})}\BibitemShut
  {NoStop}%
\bibitem [{\citenamefont {Villavicencio}\ \emph {et~al.}(2021)\citenamefont
  {Villavicencio}, \citenamefont {Cota}, \citenamefont {Rojas}, \citenamefont
  {Maytorena},\ and\ \citenamefont {Galindo}}]{PhysRevA.104.042204}%
  \BibitemOpen
  \bibfield  {author} {\bibinfo {author} {\bibfnamefont {J.}~\bibnamefont
  {Villavicencio}}, \bibinfo {author} {\bibfnamefont {E.}~\bibnamefont {Cota}},
  \bibinfo {author} {\bibfnamefont {F.}~\bibnamefont {Rojas}}, \bibinfo
  {author} {\bibfnamefont {J.~A.}\ \bibnamefont {Maytorena}},\ and\ \bibinfo
  {author} {\bibfnamefont {D.~M.}\ \bibnamefont {Galindo}},\ }\bibfield
  {title} {\bibinfo {title} {Uhlmann phase in composite systems with
  entanglement},\ }\href@noop {} {\bibfield  {journal} {\bibinfo  {journal}
  {Phys. Rev. A}\ }\textbf {\bibinfo {volume} {104}},\ \bibinfo {pages}
  {042204} (\bibinfo {year} {2021})}\BibitemShut {NoStop}%
\bibitem [{\citenamefont {Tang}\ \emph {et~al.}(2024)\citenamefont {Tang},
  \citenamefont {Hou}, \citenamefont {Zhou}, \citenamefont {Guo},\ and\
  \citenamefont {Chien}}]{PhysRevB.110.134319}%
  \BibitemOpen
  \bibfield  {author} {\bibinfo {author} {\bibfnamefont {J.-C.}\ \bibnamefont
  {Tang}}, \bibinfo {author} {\bibfnamefont {X.-Y.}\ \bibnamefont {Hou}},
  \bibinfo {author} {\bibfnamefont {Z.}~\bibnamefont {Zhou}}, \bibinfo {author}
  {\bibfnamefont {H.}~\bibnamefont {Guo}},\ and\ \bibinfo {author}
  {\bibfnamefont {C.-C.}\ \bibnamefont {Chien}},\ }\bibfield  {title} {\bibinfo
  {title} {Uhlmann quench and geometric dynamic quantum phase transition of
  mixed states},\ }\href {https://doi.org/10.1103/PhysRevB.110.134319}
  {\bibfield  {journal} {\bibinfo  {journal} {Phys. Rev. B}\ }\textbf {\bibinfo
  {volume} {110}},\ \bibinfo {pages} {134319} (\bibinfo {year}
  {2024})}\BibitemShut {NoStop}%
\bibitem [{\citenamefont {Wang}\ \emph
  {et~al.}(2025{\natexlab{a}})\citenamefont {Wang}, \citenamefont {Tang},
  \citenamefont {Hou}, \citenamefont {Guo},\ and\ \citenamefont
  {Chien}}]{98sq-16bz}%
  \BibitemOpen
  \bibfield  {author} {\bibinfo {author} {\bibfnamefont {X.}~\bibnamefont
  {Wang}}, \bibinfo {author} {\bibfnamefont {J.-C.}\ \bibnamefont {Tang}},
  \bibinfo {author} {\bibfnamefont {X.-Y.}\ \bibnamefont {Hou}}, \bibinfo
  {author} {\bibfnamefont {H.}~\bibnamefont {Guo}},\ and\ \bibinfo {author}
  {\bibfnamefont {C.-C.}\ \bibnamefont {Chien}},\ }\bibfield  {title} {\bibinfo
  {title} {Mixed-state geometric phases of coherent and squeezed spin states},\
  }\href {https://doi.org/10.1103/98sq-16bz} {\bibfield  {journal} {\bibinfo
  {journal} {Phys. Rev. B}\ }\textbf {\bibinfo {volume} {111}},\ \bibinfo
  {pages} {235450} (\bibinfo {year} {2025}{\natexlab{a}})}\BibitemShut
  {NoStop}%
\bibitem [{\citenamefont {Wang}\ \emph
  {et~al.}(2025{\natexlab{b}})\citenamefont {Wang}, \citenamefont {Hou},
  \citenamefont {He},\ and\ \citenamefont {Guo}}]{prq8-c9ns}%
  \BibitemOpen
  \bibfield  {author} {\bibinfo {author} {\bibfnamefont {X.}~\bibnamefont
  {Wang}}, \bibinfo {author} {\bibfnamefont {X.-Y.}\ \bibnamefont {Hou}},
  \bibinfo {author} {\bibfnamefont {Y.}~\bibnamefont {He}},\ and\ \bibinfo
  {author} {\bibfnamefont {H.}~\bibnamefont {Guo}},\ }\bibfield  {title}
  {\bibinfo {title} {Thermal uhlmann-chern number: Bridging pure and mixed
  states},\ }\href {https://doi.org/10.1103/prq8-c9ns} {\bibfield  {journal}
  {\bibinfo  {journal} {Phys. Rev. B}\ }\textbf {\bibinfo {volume} {112}},\
  \bibinfo {pages} {214112} (\bibinfo {year} {2025}{\natexlab{b}})}\BibitemShut
  {NoStop}%
\bibitem [{\citenamefont {Hatano}\ and\ \citenamefont
  {Nelson}(1996)}]{PhysRevLett.77.570}%
  \BibitemOpen
  \bibfield  {author} {\bibinfo {author} {\bibfnamefont {N.}~\bibnamefont
  {Hatano}}\ and\ \bibinfo {author} {\bibfnamefont {D.~R.}\ \bibnamefont
  {Nelson}},\ }\bibfield  {title} {\bibinfo {title} {Localization transitions
  in non-hermitian quantum mechanics},\ }\href
  {https://doi.org/10.1103/PhysRevLett.77.570} {\bibfield  {journal} {\bibinfo
  {journal} {Phys. Rev. Lett.}\ }\textbf {\bibinfo {volume} {77}},\ \bibinfo
  {pages} {570} (\bibinfo {year} {1996})}\BibitemShut {NoStop}%
\bibitem [{\citenamefont {Gong}\ \emph {et~al.}(2022)\citenamefont {Gong},
  \citenamefont {Bello}, \citenamefont {Malz},\ and\ \citenamefont
  {Kunst}}]{GZ22}%
  \BibitemOpen
  \bibfield  {author} {\bibinfo {author} {\bibfnamefont {Z.}~\bibnamefont
  {Gong}}, \bibinfo {author} {\bibfnamefont {M.}~\bibnamefont {Bello}},
  \bibinfo {author} {\bibfnamefont {D.}~\bibnamefont {Malz}},\ and\ \bibinfo
  {author} {\bibfnamefont {F.~K.}\ \bibnamefont {Kunst}},\ }\bibfield  {title}
  {\bibinfo {title} {Anomalous behaviors of quantum emitters in non-hermitian
  baths},\ }\href {https://doi.org/10.1103/PhysRevLett.129.223601} {\bibfield
  {journal} {\bibinfo  {journal} {Phys. Rev. Lett.}\ }\textbf {\bibinfo
  {volume} {129}},\ \bibinfo {pages} {223601} (\bibinfo {year}
  {2022})}\BibitemShut {NoStop}%
\bibitem [{\citenamefont {Roccati}\ \emph {et~al.}(2022)\citenamefont
  {Roccati}, \citenamefont {Lorenzo}, \citenamefont {Calajò}, \citenamefont
  {Palma}, \citenamefont {Carollo},\ and\ \citenamefont
  {Ciccarello}}]{Roccati_2022}%
  \BibitemOpen
  \bibfield  {author} {\bibinfo {author} {\bibfnamefont {F.}~\bibnamefont
  {Roccati}}, \bibinfo {author} {\bibfnamefont {S.}~\bibnamefont {Lorenzo}},
  \bibinfo {author} {\bibfnamefont {G.}~\bibnamefont {Calajò}}, \bibinfo
  {author} {\bibfnamefont {G.~M.}\ \bibnamefont {Palma}}, \bibinfo {author}
  {\bibfnamefont {A.}~\bibnamefont {Carollo}},\ and\ \bibinfo {author}
  {\bibfnamefont {F.}~\bibnamefont {Ciccarello}},\ }\bibfield  {title}
  {\bibinfo {title} {Exotic interactions mediated by a non-hermitian photonic
  bath},\ }\href {https://doi.org/10.1364/optica.443955} {\bibfield  {journal}
  {\bibinfo  {journal} {Optica}\ }\textbf {\bibinfo {volume} {9}},\ \bibinfo
  {pages} {565} (\bibinfo {year} {2022})}\BibitemShut {NoStop}%
\bibitem [{\citenamefont {Gong}\ \emph {et~al.}(2018)\citenamefont {Gong},
  \citenamefont {Ashida}, \citenamefont {Kawabata}, \citenamefont {Takasan},
  \citenamefont {Higashikawa},\ and\ \citenamefont {Ueda}}]{PhysRevX.8.031079}%
  \BibitemOpen
  \bibfield  {author} {\bibinfo {author} {\bibfnamefont {Z.}~\bibnamefont
  {Gong}}, \bibinfo {author} {\bibfnamefont {Y.}~\bibnamefont {Ashida}},
  \bibinfo {author} {\bibfnamefont {K.}~\bibnamefont {Kawabata}}, \bibinfo
  {author} {\bibfnamefont {K.}~\bibnamefont {Takasan}}, \bibinfo {author}
  {\bibfnamefont {S.}~\bibnamefont {Higashikawa}},\ and\ \bibinfo {author}
  {\bibfnamefont {M.}~\bibnamefont {Ueda}},\ }\bibfield  {title} {\bibinfo
  {title} {Topological phases of non-hermitian systems},\ }\href
  {https://doi.org/10.1103/PhysRevX.8.031079} {\bibfield  {journal} {\bibinfo
  {journal} {Phys. Rev. X}\ }\textbf {\bibinfo {volume} {8}},\ \bibinfo {pages}
  {031079} (\bibinfo {year} {2018})}\BibitemShut {NoStop}%
\bibitem [{\citenamefont {Kunst}\ \emph {et~al.}(2018)\citenamefont {Kunst},
  \citenamefont {Edvardsson}, \citenamefont {Budich},\ and\ \citenamefont
  {Bergholtz}}]{PhysRevLett.121.026808}%
  \BibitemOpen
  \bibfield  {author} {\bibinfo {author} {\bibfnamefont {F.~K.}\ \bibnamefont
  {Kunst}}, \bibinfo {author} {\bibfnamefont {E.}~\bibnamefont {Edvardsson}},
  \bibinfo {author} {\bibfnamefont {J.~C.}\ \bibnamefont {Budich}},\ and\
  \bibinfo {author} {\bibfnamefont {E.~J.}\ \bibnamefont {Bergholtz}},\
  }\bibfield  {title} {\bibinfo {title} {Biorthogonal bulk-boundary
  correspondence in non-hermitian systems},\ }\href
  {https://doi.org/10.1103/PhysRevLett.121.026808} {\bibfield  {journal}
  {\bibinfo  {journal} {Phys. Rev. Lett.}\ }\textbf {\bibinfo {volume} {121}},\
  \bibinfo {pages} {026808} (\bibinfo {year} {2018})}\BibitemShut {NoStop}%
\bibitem [{\citenamefont {Roccati}\ \emph {et~al.}(2024)\citenamefont
  {Roccati}, \citenamefont {Bello}, \citenamefont {Gong}, \citenamefont {Ueda},
  \citenamefont {Ciccarello}, \citenamefont {Chenu},\ and\ \citenamefont
  {Carollo}}]{Roccati2023}%
  \BibitemOpen
  \bibfield  {author} {\bibinfo {author} {\bibfnamefont {F.}~\bibnamefont
  {Roccati}}, \bibinfo {author} {\bibfnamefont {M.}~\bibnamefont {Bello}},
  \bibinfo {author} {\bibfnamefont {Z.}~\bibnamefont {Gong}}, \bibinfo {author}
  {\bibfnamefont {M.}~\bibnamefont {Ueda}}, \bibinfo {author} {\bibfnamefont
  {F.}~\bibnamefont {Ciccarello}}, \bibinfo {author} {\bibfnamefont
  {A.}~\bibnamefont {Chenu}},\ and\ \bibinfo {author} {\bibfnamefont
  {A.}~\bibnamefont {Carollo}},\ }\bibfield  {title} {\bibinfo {title}
  {Hermitian and non-hermitian topology from photon-mediated interactions},\
  }\href {https://doi.org/10.1038/s41467-024-46738-6} {\bibfield  {journal}
  {\bibinfo  {journal} {Nat. Commun.}\ }\textbf {\bibinfo {volume} {15}},\
  \bibinfo {pages} {2400} (\bibinfo {year} {2024})}\BibitemShut {NoStop}%
\bibitem [{\citenamefont {Yao}\ and\ \citenamefont
  {Wang}(2018)}]{PhysRevLett.121.086803}%
  \BibitemOpen
  \bibfield  {author} {\bibinfo {author} {\bibfnamefont {S.}~\bibnamefont
  {Yao}}\ and\ \bibinfo {author} {\bibfnamefont {Z.}~\bibnamefont {Wang}},\
  }\bibfield  {title} {\bibinfo {title} {Edge states and topological invariants
  of non-hermitian systems},\ }\href
  {https://doi.org/10.1103/PhysRevLett.121.086803} {\bibfield  {journal}
  {\bibinfo  {journal} {Phys. Rev. Lett.}\ }\textbf {\bibinfo {volume} {121}},\
  \bibinfo {pages} {086803} (\bibinfo {year} {2018})}\BibitemShut {NoStop}%
\bibitem [{\citenamefont {Li}\ \emph {et~al.}(2019)\citenamefont {Li},
  \citenamefont {Harter}, \citenamefont {Liu}, \citenamefont {de~Melo},
  \citenamefont {Joglekar},\ and\ \citenamefont {Luo}}]{Li2019NatCommun}%
  \BibitemOpen
  \bibfield  {author} {\bibinfo {author} {\bibfnamefont {J.}~\bibnamefont
  {Li}}, \bibinfo {author} {\bibfnamefont {A.~K.}\ \bibnamefont {Harter}},
  \bibinfo {author} {\bibfnamefont {J.}~\bibnamefont {Liu}}, \bibinfo {author}
  {\bibfnamefont {L.}~\bibnamefont {de~Melo}}, \bibinfo {author} {\bibfnamefont
  {Y.~N.}\ \bibnamefont {Joglekar}},\ and\ \bibinfo {author} {\bibfnamefont
  {L.}~\bibnamefont {Luo}},\ }\bibfield  {title} {\bibinfo {title} {Observation
  of parity–time symmetry breaking transitions in a dissipative floquet
  system of ultracold atoms},\ }\href
  {https://doi.org/10.1038/s41467-019-08596-1} {\bibfield  {journal} {\bibinfo
  {journal} {Nature Communications}\ }\textbf {\bibinfo {volume} {10}},\
  \bibinfo {pages} {855} (\bibinfo {year} {2019})}\BibitemShut {NoStop}%
\bibitem [{\citenamefont {Xiao}\ \emph {et~al.}(2019)\citenamefont {Xiao},
  \citenamefont {Wang}, \citenamefont {Zhan}, \citenamefont {Bian},
  \citenamefont {Kawabata}, \citenamefont {Ueda}, \citenamefont {Yi},\ and\
  \citenamefont {Xue}}]{PhysRevLett.123.230401}%
  \BibitemOpen
  \bibfield  {author} {\bibinfo {author} {\bibfnamefont {L.}~\bibnamefont
  {Xiao}}, \bibinfo {author} {\bibfnamefont {K.}~\bibnamefont {Wang}}, \bibinfo
  {author} {\bibfnamefont {X.}~\bibnamefont {Zhan}}, \bibinfo {author}
  {\bibfnamefont {Z.}~\bibnamefont {Bian}}, \bibinfo {author} {\bibfnamefont
  {K.}~\bibnamefont {Kawabata}}, \bibinfo {author} {\bibfnamefont
  {M.}~\bibnamefont {Ueda}}, \bibinfo {author} {\bibfnamefont {W.}~\bibnamefont
  {Yi}},\ and\ \bibinfo {author} {\bibfnamefont {P.}~\bibnamefont {Xue}},\
  }\bibfield  {title} {\bibinfo {title} {Observation of critical phenomena in
  parity-time-symmetric quantum dynamics},\ }\href
  {https://doi.org/10.1103/PhysRevLett.123.230401} {\bibfield  {journal}
  {\bibinfo  {journal} {Phys. Rev. Lett.}\ }\textbf {\bibinfo {volume} {123}},\
  \bibinfo {pages} {230401} (\bibinfo {year} {2019})}\BibitemShut {NoStop}%
\bibitem [{\citenamefont {Wu}\ \emph {et~al.}(2019)\citenamefont {Wu},
  \citenamefont {Liu}, \citenamefont {Geng}, \citenamefont {Song},
  \citenamefont {Ye}, \citenamefont {Duan}, \citenamefont {Rong},\ and\
  \citenamefont {Du}}]{doi:10.1126/science.aaw8205}%
  \BibitemOpen
  \bibfield  {author} {\bibinfo {author} {\bibfnamefont {Y.}~\bibnamefont
  {Wu}}, \bibinfo {author} {\bibfnamefont {W.}~\bibnamefont {Liu}}, \bibinfo
  {author} {\bibfnamefont {J.}~\bibnamefont {Geng}}, \bibinfo {author}
  {\bibfnamefont {X.}~\bibnamefont {Song}}, \bibinfo {author} {\bibfnamefont
  {X.}~\bibnamefont {Ye}}, \bibinfo {author} {\bibfnamefont {C.-K.}\
  \bibnamefont {Duan}}, \bibinfo {author} {\bibfnamefont {X.}~\bibnamefont
  {Rong}},\ and\ \bibinfo {author} {\bibfnamefont {J.}~\bibnamefont {Du}},\
  }\bibfield  {title} {\bibinfo {title} {Observation of parity-time symmetry
  breaking in a single-spin system},\ }\href@noop {} {\bibfield  {journal}
  {\bibinfo  {journal} {Science}\ }\textbf {\bibinfo {volume} {364}},\ \bibinfo
  {pages} {878} (\bibinfo {year} {2019})}\BibitemShut {NoStop}%
\bibitem [{\citenamefont {Ding}\ \emph {et~al.}(2021)\citenamefont {Ding},
  \citenamefont {Shi}, \citenamefont {Zhang}, \citenamefont {Shen},
  \citenamefont {Zhang},\ and\ \citenamefont {Zhang}}]{PhysRevLett.126.083604}%
  \BibitemOpen
  \bibfield  {author} {\bibinfo {author} {\bibfnamefont {L.}~\bibnamefont
  {Ding}}, \bibinfo {author} {\bibfnamefont {K.}~\bibnamefont {Shi}}, \bibinfo
  {author} {\bibfnamefont {Q.}~\bibnamefont {Zhang}}, \bibinfo {author}
  {\bibfnamefont {D.}~\bibnamefont {Shen}}, \bibinfo {author} {\bibfnamefont
  {X.}~\bibnamefont {Zhang}},\ and\ \bibinfo {author} {\bibfnamefont
  {W.}~\bibnamefont {Zhang}},\ }\bibfield  {title} {\bibinfo {title}
  {Experimental determination of $\mathcal{P}\mathcal{T}$-symmetric exceptional
  points in a single trapped ion},\ }\href
  {https://doi.org/10.1103/PhysRevLett.126.083604} {\bibfield  {journal}
  {\bibinfo  {journal} {Phys. Rev. Lett.}\ }\textbf {\bibinfo {volume} {126}},\
  \bibinfo {pages} {083604} (\bibinfo {year} {2021})}\BibitemShut {NoStop}%
\bibitem [{\citenamefont {Xiao}\ \emph {et~al.}(2020)\citenamefont {Xiao},
  \citenamefont {Deng}, \citenamefont {Wang}, \citenamefont {Yi},\ and\
  \citenamefont {Xue}}]{Xiao2020NatPhys}%
  \BibitemOpen
  \bibfield  {author} {\bibinfo {author} {\bibfnamefont {L.}~\bibnamefont
  {Xiao}}, \bibinfo {author} {\bibfnamefont {T.}~\bibnamefont {Deng}}, \bibinfo
  {author} {\bibfnamefont {K.}~\bibnamefont {Wang}}, \bibinfo {author}
  {\bibfnamefont {W.}~\bibnamefont {Yi}},\ and\ \bibinfo {author}
  {\bibfnamefont {P.}~\bibnamefont {Xue}},\ }\bibfield  {title} {\bibinfo
  {title} {Non-hermitian bulk–boundary correspondence in quantum dynamics},\
  }\href {https://doi.org/10.1038/s41567-020-0836-6} {\bibfield  {journal}
  {\bibinfo  {journal} {Nature Physics}\ }\textbf {\bibinfo {volume} {16}},\
  \bibinfo {pages} {761} (\bibinfo {year} {2020})}\BibitemShut {NoStop}%
\bibitem [{\citenamefont {Zhang}\ \emph {et~al.}(2020)\citenamefont {Zhang},
  \citenamefont {Yang},\ and\ \citenamefont {Fang}}]{PhysRevLett.125.126402}%
  \BibitemOpen
  \bibfield  {author} {\bibinfo {author} {\bibfnamefont {K.}~\bibnamefont
  {Zhang}}, \bibinfo {author} {\bibfnamefont {Z.}~\bibnamefont {Yang}},\ and\
  \bibinfo {author} {\bibfnamefont {C.}~\bibnamefont {Fang}},\ }\bibfield
  {title} {\bibinfo {title} {Correspondence between winding numbers and skin
  modes in non-hermitian systems},\ }\href@noop {} {\bibfield  {journal}
  {\bibinfo  {journal} {Phys. Rev. Lett.}\ }\textbf {\bibinfo {volume} {125}},\
  \bibinfo {pages} {126402} (\bibinfo {year} {2020})}\BibitemShut {NoStop}%
\bibitem [{\citenamefont {Deng}\ and\ \citenamefont {Cheng}(2025)}]{kcls-cn82}%
  \BibitemOpen
  \bibfield  {author} {\bibinfo {author} {\bibfnamefont {K.}~\bibnamefont
  {Deng}}\ and\ \bibinfo {author} {\bibfnamefont {R.}~\bibnamefont {Cheng}},\
  }\bibfield  {title} {\bibinfo {title} {Emerging $(2+1)\mathrm{D}$
  electrodynamics and topological instanton in pseudo-hermitian two-level
  systems},\ }\href {https://doi.org/10.1103/kcls-cn82} {\bibfield  {journal}
  {\bibinfo  {journal} {Phys. Rev. Lett.}\ }\textbf {\bibinfo {volume} {135}},\
  \bibinfo {pages} {126609} (\bibinfo {year} {2025})}\BibitemShut {NoStop}%
\bibitem [{\citenamefont {Gong}\ and\ \citenamefont {Wang}(2013)}]{JPAGW13}%
  \BibitemOpen
  \bibfield  {author} {\bibinfo {author} {\bibfnamefont {J.}~\bibnamefont
  {Gong}}\ and\ \bibinfo {author} {\bibfnamefont {Q.-h.}\ \bibnamefont
  {Wang}},\ }\bibfield  {title} {\bibinfo {title} {Time-dependent
  $\mathcal{PT}$-symmetric quantum mechanics},\ }\href@noop {} {\bibfield
  {journal} {\bibinfo  {journal} {J. Phys. A: Math. Theor.}\ }\textbf {\bibinfo
  {volume} {46}},\ \bibinfo {pages} {485302} (\bibinfo {year}
  {2013})}\BibitemShut {NoStop}%
\bibitem [{\citenamefont {Zhang}\ \emph {et~al.}(2019)\citenamefont {Zhang},
  \citenamefont {Wang},\ and\ \citenamefont {Gong}}]{PhysRevA.99.042104}%
  \BibitemOpen
  \bibfield  {author} {\bibinfo {author} {\bibfnamefont {D.-J.}\ \bibnamefont
  {Zhang}}, \bibinfo {author} {\bibfnamefont {Q.-h.}\ \bibnamefont {Wang}},\
  and\ \bibinfo {author} {\bibfnamefont {J.}~\bibnamefont {Gong}},\ }\bibfield
  {title} {\bibinfo {title} {Quantum geometric tensor in
  $\mathcal{PT}$-symmetric quantum mechanics},\ }\href
  {https://doi.org/10.1103/PhysRevA.99.042104} {\bibfield  {journal} {\bibinfo
  {journal} {Phys. Rev. A}\ }\textbf {\bibinfo {volume} {99}},\ \bibinfo
  {pages} {042104} (\bibinfo {year} {2019})}\BibitemShut {NoStop}%
\bibitem [{\citenamefont {Yang}\ \emph {et~al.}(2026)\citenamefont {Yang},
  \citenamefont {Han}, \citenamefont {Ning}, \citenamefont {Wu}, \citenamefont
  {Yang},\ and\ \citenamefont {Zheng}}]{Yang2026MixedStateTopology}%
  \BibitemOpen
  \bibfield  {author} {\bibinfo {author} {\bibfnamefont {S.-B.}\ \bibnamefont
  {Yang}}, \bibinfo {author} {\bibfnamefont {P.-R.}\ \bibnamefont {Han}},
  \bibinfo {author} {\bibfnamefont {W.}~\bibnamefont {Ning}}, \bibinfo {author}
  {\bibfnamefont {F.}~\bibnamefont {Wu}}, \bibinfo {author} {\bibfnamefont
  {Z.-B.}\ \bibnamefont {Yang}},\ and\ \bibinfo {author} {\bibfnamefont
  {S.-B.}\ \bibnamefont {Zheng}},\ }\href {https://arxiv.org/abs/2602.10831}
  {\bibinfo {title} {Mixed-state topology in non-hermitian systems}} (\bibinfo
  {year} {2026}),\ \Eprint {https://arxiv.org/abs/2602.10831} {arXiv:2602.10831
  [quant-ph]} \BibitemShut {NoStop}%
\bibitem [{\citenamefont {Pauli}(1943)}]{RevModPhys.15.175}%
  \BibitemOpen
  \bibfield  {author} {\bibinfo {author} {\bibfnamefont {W.}~\bibnamefont
  {Pauli}},\ }\bibfield  {title} {\bibinfo {title} {On dirac's new method of
  field quantization},\ }\href {https://doi.org/10.1103/RevModPhys.15.175}
  {\bibfield  {journal} {\bibinfo  {journal} {Rev. Mod. Phys.}\ }\textbf
  {\bibinfo {volume} {15}},\ \bibinfo {pages} {175} (\bibinfo {year}
  {1943})}\BibitemShut {NoStop}%
\bibitem [{\citenamefont {Bender}\ and\ \citenamefont
  {Boettcher}(1998)}]{PhysRevLett.80.5243}%
  \BibitemOpen
  \bibfield  {author} {\bibinfo {author} {\bibfnamefont {C.~M.}\ \bibnamefont
  {Bender}}\ and\ \bibinfo {author} {\bibfnamefont {S.}~\bibnamefont
  {Boettcher}},\ }\bibfield  {title} {\bibinfo {title} {Real spectra in
  non-hermitian hamiltonians having pt symmetry},\ }\href@noop {} {\bibfield
  {journal} {\bibinfo  {journal} {Phys. Rev. Lett.}\ }\textbf {\bibinfo
  {volume} {80}},\ \bibinfo {pages} {5243} (\bibinfo {year}
  {1998})}\BibitemShut {NoStop}%
\bibitem [{\citenamefont
  {Mostafazadeh}(2002{\natexlab{a}})}]{Mostafazadeh2002I}%
  \BibitemOpen
  \bibfield  {author} {\bibinfo {author} {\bibfnamefont {A.}~\bibnamefont
  {Mostafazadeh}},\ }\bibfield  {title} {\bibinfo {title} {Pseudo-hermiticity
  versus pt symmetry: The necessary condition for the reality of the spectrum
  of a non-hermitian hamiltonian},\ }\href {https://doi.org/10.1063/1.1418246}
  {\bibfield  {journal} {\bibinfo  {journal} {J. Math. Phys.}\ }\textbf
  {\bibinfo {volume} {43}},\ \bibinfo {pages} {205} (\bibinfo {year}
  {2002}{\natexlab{a}})}\BibitemShut {NoStop}%
\bibitem [{\citenamefont
  {Mostafazadeh}(2002{\natexlab{b}})}]{Mostafazadeh2002II}%
  \BibitemOpen
  \bibfield  {author} {\bibinfo {author} {\bibfnamefont {A.}~\bibnamefont
  {Mostafazadeh}},\ }\bibfield  {title} {\bibinfo {title} {Pseudo-hermiticity
  versus pt symmetry ii: A complete characterization of non-hermitian
  hamiltonians with a real spectrum},\ }\href
  {https://doi.org/10.1063/1.1461427} {\bibfield  {journal} {\bibinfo
  {journal} {J. Math. Phys.}\ }\textbf {\bibinfo {volume} {43}},\ \bibinfo
  {pages} {2814} (\bibinfo {year} {2002}{\natexlab{b}})}\BibitemShut {NoStop}%
\bibitem [{\citenamefont
  {Mostafazadeh}(2002{\natexlab{c}})}]{Mostafazadeh2002III}%
  \BibitemOpen
  \bibfield  {author} {\bibinfo {author} {\bibfnamefont {A.}~\bibnamefont
  {Mostafazadeh}},\ }\bibfield  {title} {\bibinfo {title} {Pseudo-hermiticity
  versus pt symmetry iii: Equivalence of pseudo-hermiticity and the presence of
  antilinear symmetries},\ }\href {https://doi.org/10.1063/1.1489072}
  {\bibfield  {journal} {\bibinfo  {journal} {J. Math. Phys.}\ }\textbf
  {\bibinfo {volume} {43}},\ \bibinfo {pages} {3944} (\bibinfo {year}
  {2002}{\natexlab{c}})}\BibitemShut {NoStop}%
\bibitem [{\citenamefont {Giorgi}(2010)}]{PhysRevB.82.052404}%
  \BibitemOpen
  \bibfield  {author} {\bibinfo {author} {\bibfnamefont {G.~L.}\ \bibnamefont
  {Giorgi}},\ }\bibfield  {title} {\bibinfo {title} {Spontaneous
  $\mathcal{P}\mathcal{T}$ symmetry breaking and quantum phase transitions in
  dimerized spin chains},\ }\href {https://doi.org/10.1103/PhysRevB.82.052404}
  {\bibfield  {journal} {\bibinfo  {journal} {Phys. Rev. B}\ }\textbf {\bibinfo
  {volume} {82}},\ \bibinfo {pages} {052404} (\bibinfo {year}
  {2010})}\BibitemShut {NoStop}%
\bibitem [{\citenamefont {Hang}\ \emph {et~al.}(2013)\citenamefont {Hang},
  \citenamefont {Huang},\ and\ \citenamefont
  {Konotop}}]{PhysRevLett.110.083604}%
  \BibitemOpen
  \bibfield  {author} {\bibinfo {author} {\bibfnamefont {C.}~\bibnamefont
  {Hang}}, \bibinfo {author} {\bibfnamefont {G.}~\bibnamefont {Huang}},\ and\
  \bibinfo {author} {\bibfnamefont {V.~V.}\ \bibnamefont {Konotop}},\
  }\bibfield  {title} {\bibinfo {title} {$\mathcal{P}\mathcal{T}$ symmetry with
  a system of three-level atoms},\ }\href
  {https://doi.org/10.1103/PhysRevLett.110.083604} {\bibfield  {journal}
  {\bibinfo  {journal} {Phys. Rev. Lett.}\ }\textbf {\bibinfo {volume} {110}},\
  \bibinfo {pages} {083604} (\bibinfo {year} {2013})}\BibitemShut {NoStop}%
\bibitem [{\citenamefont {Korff}\ and\ \citenamefont
  {Weston}(2007)}]{Korff_2007}%
  \BibitemOpen
  \bibfield  {author} {\bibinfo {author} {\bibfnamefont {C.}~\bibnamefont
  {Korff}}\ and\ \bibinfo {author} {\bibfnamefont {R.}~\bibnamefont {Weston}},\
  }\bibfield  {title} {\bibinfo {title} {Pt symmetry on the lattice: the
  quantum group invariant xxz spin chain},\ }\href
  {https://doi.org/10.1088/1751-8113/40/30/016} {\bibfield  {journal} {\bibinfo
   {journal} {J. Phys. A: Math. Theor.}\ }\textbf {\bibinfo {volume} {40}},\
  \bibinfo {pages} {8845} (\bibinfo {year} {2007})}\BibitemShut {NoStop}%
\bibitem [{\citenamefont {Korff}(2008)}]{Korff_2008}%
  \BibitemOpen
  \bibfield  {author} {\bibinfo {author} {\bibfnamefont {C.}~\bibnamefont
  {Korff}},\ }\bibfield  {title} {\bibinfo {title} {Pt symmetry of the
  non-hermitian xx spin-chain: non-local bulk interaction from complex boundary
  fields},\ }\href {https://doi.org/10.1088/1751-8113/41/29/295206} {\bibfield
  {journal} {\bibinfo  {journal} {J. Phys. A: Math. Theor.}\ }\textbf {\bibinfo
  {volume} {41}},\ \bibinfo {pages} {295206} (\bibinfo {year}
  {2008})}\BibitemShut {NoStop}%
\bibitem [{\citenamefont {Konotop}\ \emph {et~al.}(2016)\citenamefont
  {Konotop}, \citenamefont {Yang},\ and\ \citenamefont
  {Zezyulin}}]{RevModPhys.88.035002}%
  \BibitemOpen
  \bibfield  {author} {\bibinfo {author} {\bibfnamefont {V.~V.}\ \bibnamefont
  {Konotop}}, \bibinfo {author} {\bibfnamefont {J.}~\bibnamefont {Yang}},\ and\
  \bibinfo {author} {\bibfnamefont {D.~A.}\ \bibnamefont {Zezyulin}},\
  }\bibfield  {title} {\bibinfo {title} {Nonlinear waves in
  $\mathcal{PT}$-symmetric systems},\ }\href
  {https://doi.org/10.1103/RevModPhys.88.035002} {\bibfield  {journal}
  {\bibinfo  {journal} {Rev. Mod. Phys.}\ }\textbf {\bibinfo {volume} {88}},\
  \bibinfo {pages} {035002} (\bibinfo {year} {2016})}\BibitemShut {NoStop}%
\bibitem [{\citenamefont {Bender}\ \emph {et~al.}(1999)\citenamefont {Bender},
  \citenamefont {Boettcher},\ and\ \citenamefont {Meisinger}}]{Bender1999}%
  \BibitemOpen
  \bibfield  {author} {\bibinfo {author} {\bibfnamefont {C.~M.}\ \bibnamefont
  {Bender}}, \bibinfo {author} {\bibfnamefont {S.}~\bibnamefont {Boettcher}},\
  and\ \bibinfo {author} {\bibfnamefont {P.~N.}\ \bibnamefont {Meisinger}},\
  }\bibfield  {title} {\bibinfo {title} {$\mathcal{P}\mathcal{T}$-symmetric
  quantum mechanics},\ }\href {https://doi.org/10.1063/1.532860} {\bibfield
  {journal} {\bibinfo  {journal} {Journal of Mathematical Physics}\ }\textbf
  {\bibinfo {volume} {40}},\ \bibinfo {pages} {2201–2229} (\bibinfo {year}
  {1999})}\BibitemShut {NoStop}%
\bibitem [{\citenamefont {Cham}(2015)}]{Cham_2015}%
  \BibitemOpen
  \bibfield  {author} {\bibinfo {author} {\bibfnamefont {J.}~\bibnamefont
  {Cham}},\ }\bibfield  {title} {\bibinfo {title} {Top 10 physics discoveries
  of the last 10 years},\ }\href {https://doi.org/10.1038/nphys3500} {\bibfield
   {journal} {\bibinfo  {journal} {Nature Physics}\ }\textbf {\bibinfo {volume}
  {11}},\ \bibinfo {pages} {799–799} (\bibinfo {year} {2015})}\BibitemShut
  {NoStop}%
\bibitem [{\citenamefont {Feng}\ \emph {et~al.}(2017)\citenamefont {Feng},
  \citenamefont {El-Ganainy},\ and\ \citenamefont {Ge}}]{Feng_2017}%
  \BibitemOpen
  \bibfield  {author} {\bibinfo {author} {\bibfnamefont {L.}~\bibnamefont
  {Feng}}, \bibinfo {author} {\bibfnamefont {R.}~\bibnamefont {El-Ganainy}},\
  and\ \bibinfo {author} {\bibfnamefont {L.}~\bibnamefont {Ge}},\ }\bibfield
  {title} {\bibinfo {title} {Non-hermitian photonics based on parity–time
  symmetry},\ }\href {https://doi.org/10.1038/s41566-017-0031-1} {\bibfield
  {journal} {\bibinfo  {journal} {Nature Photonics}\ }\textbf {\bibinfo
  {volume} {11}},\ \bibinfo {pages} {752–762} (\bibinfo {year}
  {2017})}\BibitemShut {NoStop}%
\bibitem [{\citenamefont {MOSTAFAZADEH}(2010)}]{doi:10.1142/S0219887810004816}%
  \BibitemOpen
  \bibfield  {author} {\bibinfo {author} {\bibfnamefont {A.}~\bibnamefont
  {MOSTAFAZADEH}},\ }\bibfield  {title} {\bibinfo {title} {Pseudo-hermitian
  representation of quantum mechanics},\ }\href@noop {} {\bibfield  {journal}
  {\bibinfo  {journal} {International Journal of Geometric Methods in Modern
  Physics}\ }\textbf {\bibinfo {volume} {07}},\ \bibinfo {pages} {1191}
  (\bibinfo {year} {2010})}\BibitemShut {NoStop}%
\bibitem [{\citenamefont {Das}(2011)}]{Das_2011}%
  \BibitemOpen
  \bibfield  {author} {\bibinfo {author} {\bibfnamefont {A.}~\bibnamefont
  {Das}},\ }\bibfield  {title} {\bibinfo {title} {Pseudo-hermitian quantum
  mechanics},\ }\href@noop {} {\bibfield  {journal} {\bibinfo  {journal}
  {Journal of Physics: Conference Series}\ }\textbf {\bibinfo {volume} {287}},\
  \bibinfo {pages} {012002} (\bibinfo {year} {2011})}\BibitemShut {NoStop}%
\bibitem [{\citenamefont {Chru\'{s}ci\'{n}ski}\ and\ \citenamefont
  {Jamio{\l}kowski}(2004)}]{ChruscinskiJamiolkowski2004}%
  \BibitemOpen
  \bibfield  {author} {\bibinfo {author} {\bibfnamefont {D.}~\bibnamefont
  {Chru\'{s}ci\'{n}ski}}\ and\ \bibinfo {author} {\bibfnamefont
  {A.}~\bibnamefont {Jamio{\l}kowski}},\ }\href
  {https://doi.org/10.1007/978-0-8176-8176-0} {\emph {\bibinfo {title}
  {Geometric Phases in Classical and Quantum Mechanics}}},\ \bibinfo {series}
  {Progress in Mathematical Physics}, Vol.~\bibinfo {volume} {36}\ (\bibinfo
  {publisher} {Birkh\"{a}user Boston, MA},\ \bibinfo {year} {2004})\BibitemShut
  {NoStop}%
\bibitem [{\citenamefont {Wang}\ \emph {et~al.}(2010)\citenamefont {Wang},
  \citenamefont {Chia},\ and\ \citenamefont {Zhang}}]{Wang_2010}%
  \BibitemOpen
  \bibfield  {author} {\bibinfo {author} {\bibfnamefont {Q.-h.}\ \bibnamefont
  {Wang}}, \bibinfo {author} {\bibfnamefont {S.-z.}\ \bibnamefont {Chia}},\
  and\ \bibinfo {author} {\bibfnamefont {J.-h.}\ \bibnamefont {Zhang}},\
  }\bibfield  {title} {\bibinfo {title} {symmetry as a generalization of
  hermiticity},\ }\href@noop {} {\bibfield  {journal} {\bibinfo  {journal}
  {Journal of Physics A: Mathematical and Theoretical}\ }\textbf {\bibinfo
  {volume} {43}},\ \bibinfo {pages} {295301} (\bibinfo {year}
  {2010})}\BibitemShut {NoStop}%
\bibitem [{\citenamefont {Bohm}\ \emph {et~al.}(2003)\citenamefont {Bohm},
  \citenamefont {Mostafazadeh}, \citenamefont {Koizumi}, \citenamefont {Niu},\
  and\ \citenamefont {Zwanziger}}]{Bohm2003GeometricPhase}%
  \BibitemOpen
  \bibfield  {author} {\bibinfo {author} {\bibfnamefont {A.}~\bibnamefont
  {Bohm}}, \bibinfo {author} {\bibfnamefont {A.}~\bibnamefont {Mostafazadeh}},
  \bibinfo {author} {\bibfnamefont {H.}~\bibnamefont {Koizumi}}, \bibinfo
  {author} {\bibfnamefont {Q.}~\bibnamefont {Niu}},\ and\ \bibinfo {author}
  {\bibfnamefont {J.}~\bibnamefont {Zwanziger}},\ }\href
  {https://doi.org/10.1007/978-3-662-10333-3} {\emph {\bibinfo {title} {The
  Geometric Phase in Quantum Systems: Foundations, Mathematical Concepts, and
  Applications in Molecular and Condensed Matter Physics}}}\ (\bibinfo
  {publisher} {Springer},\ \bibinfo {address} {Berlin, Heidelberg},\ \bibinfo
  {year} {2003})\BibitemShut {NoStop}%
\bibitem [{\citenamefont {Heyl}(2018)}]{Heyl_2018}%
  \BibitemOpen
  \bibfield  {author} {\bibinfo {author} {\bibfnamefont {M.}~\bibnamefont
  {Heyl}},\ }\bibfield  {title} {\bibinfo {title} {Dynamical quantum phase
  transitions: a review},\ }\href@noop {} {\bibfield  {journal} {\bibinfo
  {journal} {Reports on Progress in Physics}\ }\textbf {\bibinfo {volume}
  {81}},\ \bibinfo {pages} {054001} (\bibinfo {year} {2018})}\BibitemShut
  {NoStop}%
\bibitem [{\citenamefont {Mondal}\ and\ \citenamefont
  {Nag}(2023)}]{PhysRevB.107.184311}%
  \BibitemOpen
  \bibfield  {author} {\bibinfo {author} {\bibfnamefont {D.}~\bibnamefont
  {Mondal}}\ and\ \bibinfo {author} {\bibfnamefont {T.}~\bibnamefont {Nag}},\
  }\bibfield  {title} {\bibinfo {title} {Finite-temperature dynamical quantum
  phase transition in a non-hermitian system},\ }\href@noop {} {\bibfield
  {journal} {\bibinfo  {journal} {Phys. Rev. B}\ }\textbf {\bibinfo {volume}
  {107}},\ \bibinfo {pages} {184311} (\bibinfo {year} {2023})}\BibitemShut
  {NoStop}%
\bibitem [{\citenamefont {Viyuela}\ \emph {et~al.}(2018)\citenamefont
  {Viyuela}, \citenamefont {Rivas}, \citenamefont {Gasparinetti}, \citenamefont
  {Wallraff}, \citenamefont {Filipp},\ and\ \citenamefont
  {Martin-Delgado}}]{Viyuela2018TopologicalUhlmann}%
  \BibitemOpen
  \bibfield  {author} {\bibinfo {author} {\bibfnamefont {O.}~\bibnamefont
  {Viyuela}}, \bibinfo {author} {\bibfnamefont {A.}~\bibnamefont {Rivas}},
  \bibinfo {author} {\bibfnamefont {S.}~\bibnamefont {Gasparinetti}}, \bibinfo
  {author} {\bibfnamefont {A.}~\bibnamefont {Wallraff}}, \bibinfo {author}
  {\bibfnamefont {S.}~\bibnamefont {Filipp}},\ and\ \bibinfo {author}
  {\bibfnamefont {M.~A.}\ \bibnamefont {Martin-Delgado}},\ }\bibfield  {title}
  {\bibinfo {title} {Observation of topological uhlmann phases with
  superconducting qubits},\ }\href {https://doi.org/10.1038/s41534-017-0056-9}
  {\bibfield  {journal} {\bibinfo  {journal} {npj Quantum Information}\
  }\textbf {\bibinfo {volume} {4}},\ \bibinfo {pages} {10} (\bibinfo {year}
  {2018})}\BibitemShut {NoStop}%
\bibitem [{\citenamefont {Mastandrea}\ \emph {et~al.}(2026)\citenamefont
  {Mastandrea}, \citenamefont {Iancu}, \citenamefont {Guo},\ and\ \citenamefont
  {Chien}}]{Mastandrea_2026}%
  \BibitemOpen
  \bibfield  {author} {\bibinfo {author} {\bibfnamefont {C.}~\bibnamefont
  {Mastandrea}}, \bibinfo {author} {\bibfnamefont {C.}~\bibnamefont {Iancu}},
  \bibinfo {author} {\bibfnamefont {H.}~\bibnamefont {Guo}},\ and\ \bibinfo
  {author} {\bibfnamefont {C.-C.}\ \bibnamefont {Chien}},\ }\bibfield  {title}
  {\bibinfo {title} {Intermediate-temperature topological uhlmann phase on ibm
  quantum computers},\ }\href@noop {} {\bibfield  {journal} {\bibinfo
  {journal} {Quantum Science and Technology}\ }\textbf {\bibinfo {volume}
  {11}},\ \bibinfo {pages} {015033} (\bibinfo {year} {2026})}\BibitemShut
  {NoStop}%
\bibitem [{\citenamefont {Mostafazadeh}(2003)}]{Hermiticity2003}%
  \BibitemOpen
  \bibfield  {author} {\bibinfo {author} {\bibfnamefont {A.}~\bibnamefont
  {Mostafazadeh}},\ }\bibfield  {title} {\bibinfo {title} {Exact {PT}-symmetry
  is equivalent to hermiticity},\ }\href
  {https://doi.org/10.1088/0305-4470/36/25/312} {\bibfield  {journal} {\bibinfo
   {journal} {J. Phys. A: Math. Gen.}\ }\textbf {\bibinfo {volume} {36}},\
  \bibinfo {pages} {7081} (\bibinfo {year} {2003})}\BibitemShut {NoStop}%
\end{thebibliography}%
\end{document}